\documentclass[notitlepage,twocolumn,letterpaper,natbib,aps,prd,amsmath,amsfonts,nofootinbib,preprintnumbers,superscriptaddress,secnumarabic,groupedaddress]{revtex4-1}
\pdfoutput=1
\usepackage{amssymb,amsmath,latexsym,mathrsfs}
\usepackage{url}
\usepackage{enumitem}
\usepackage{graphicx}
\usepackage{booktabs}
\usepackage[usenames,dvipsnames]{color}
\usepackage[breaklinks,colorlinks,urlcolor=blue,citecolor=blue,linkcolor=magenta]{hyperref}
\usepackage{multirow}
\usepackage{float}
\usepackage{cases}
\usepackage{blindtext}
\usepackage{pifont}
\usepackage{hhline}
\usepackage{slashed}
\usepackage{array}
\usepackage{orcidlink}
\usepackage{footmisc}
\usepackage{soul}

\newcommand{\modific}[1]{{#1}}

\usepackage{xcolor}
\definecolor{linkcolor}{rgb}{0.0, 0.47, 0.75}
\definecolor{citecolor}{rgb}{1.0, 0.5, 0.0}
\hypersetup{
  linkcolor  = linkcolor,
  citecolor  = linkcolor,
  urlcolor   = linkcolor,
  colorlinks = true
}

\begin{document}

\title{Keeping It Renormalizable:\\ Minimal Baryogenesis induced Asymmetric Dark Matter}

\author{Miguel Escudero Abenza \orcidlink{0000-0002-4487-8742}}
\email{miguel.escudero@cern.ch}
\affiliation{Theoretical Physics Department, CERN, 1211 Geneva 23, Switzerland}

\author{Thomas Hambye \orcidlink{0000-0003-4381-3119}}
\email{thomas.hambye@ulb.be}
\affiliation{Service de Physique Th\'eorique, Universit\'e Libre de Bruxelles, Boulevard du Triomphe, CP225, 1050 Brussels, Belgium}

\author{Chandan Hati \orcidlink{0000-0003-4033-0082}}
\email{chandan@ific.uv.es}
\affiliation{Instituto de F\'{\i}sica Corpuscular (IFIC), CSIC‐Universitat de Val\`{e}ncia, Spain}

\date{\today}

\preprint{CERN-TH-2025-234, ULB-TH/25-08}

\begin{abstract}
\noindent Many asymmetric dark matter scenarios have been proposed to date. Among them, perhaps the most motivated ones are those in which the dark matter asymmetry is induced from the baryon/lepton asymmetries via chemical equilibration without any new sources of CP violation. However, most of the models put forward along these lines have been excluded by now and/or are based on complicated setups. In this letter, we present a new, simple, and viable scenario. It assumes only two new fields: a scalar singlet and an inert scalar doublet, and is based only on \textit{renormalizable} interactions, that slowly generate the dark matter asymmetry from the Standard Model Higgs asymmetry. The successful asymmetry transfer requires small (but technically natural) couplings which in turn make the model CP conserving in practice. The model allows for the direct detection of dark matter in the upcoming generation of experiments, and the inert doublet is predicted to be light enough to be potentially produced and observed at the LHC and future colliders, $46\,{\rm GeV}<m_{H'}<580\,{\rm GeV}$. Quite remarkably, the mass and lifetime of the inert doublet may be inferred from direct detection experiments, indicating long-lived compressed electroweak states that can be searched for at the HL-LHC.
\end{abstract}

\maketitle


\section{Introduction and Motivation} 

Ordinary matter in the Universe is asymmetric: we live in a Universe where structures are made of baryons with essentially no antibaryons. The baryons that form the Universe today are those generally expected to have survived the matter-antimatter annihilation in the early Universe at a temperature $T\simeq 40\,{\rm MeV}$~\cite{Kolb:1990vq,Rubakov:2017xzr}. Given this situation for the baryons, one wonders if the dark matter (DM) relic density today could also stem from a particle-antiparticle dark matter asymmetry created prior to a dark matter annihilation, and if the two asymmetries are connected in any way.

\begin{figure*}[!t]
\centering
\includegraphics[width=0.98\textwidth]{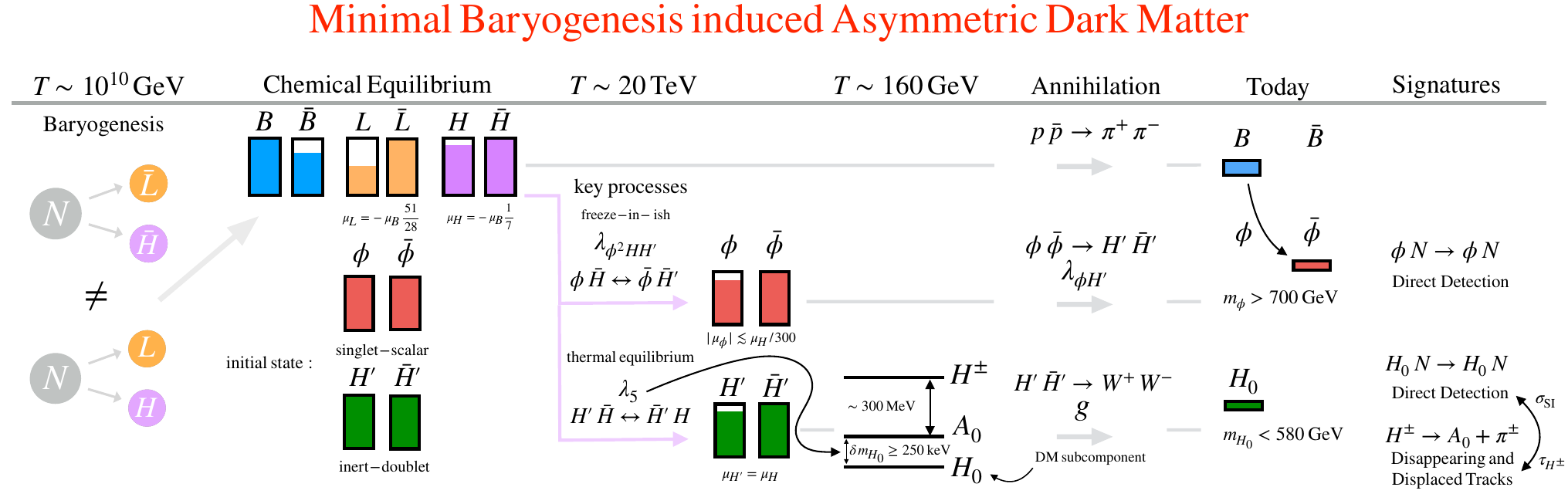}
\vspace{-0.2cm}
\caption{Sketch of our scenario for minimal baryogenesis induced asymmetric dark matter, based on two new fields $H'$ (a new inert Higgs doublet), and $\phi$ (a complex singlet scalar). Starting from a baryon asymmetry (generated by e.g. thermal leptogenesis) chemical equilibrium implies a SM Higgs doublet asymmetry at $T>T_{\rm EW}$, see footnote~\ref{footimportant}. The $\lambda_{\phi^2HH'} \sim 10^{-6}$ coupling slowly creates the $\phi$ asymmetry from the SM Higgs asymmetry, thus connecting it to the baryon asymmetry of the Universe. $\lambda_5$ triggers interactions that thermalize, leading to $\mu_{H'} = \mu_H$ and provides a mass splitting between $A_0$ and $H_0$ that avoids $Z$-mediated direct detection from the inert doublet. After annihilations, we end up with a two-component dark matter scenario: a dominant asymmetric $\phi$ population and a sub-leading symmetric $H_0$ population. $\Omega_{H_0} h^2 = 0.12\,{\rm } (m_{H_0}/580\,{\rm GeV})^2$ and hence the lighter $H_0$ the more asymmetric is dark matter. 
}
\label{fig:Mechanism_sketch}
\end{figure*}

Among the many proposed asymmetric dark matter creation scenarios (see~\cite{Cirelli:2024ssz,Zurek:2013wia,Petraki:2013wwa,Boucenna:2013wba,Davoudiasl:2012uw} for reviews), one particularly simple and motivated general class of models assumes that the dark matter asymmetry is induced via chemical equilibration from the primordial asymmetries created during baryogenesis.
This is economical because it does not require an all-encompassing dark sector with its own CP violation to create the dark matter asymmetry. Here, the CP-violation assumed reduces to the one that is at the origin of baryogenesis. For instance, within the leptogenesis mechanism~\cite{Fukugita:1986hr}, which, as of today, provides the most straightforward and motivated explanation for baryogenesis, lepton and Higgs asymmetries are generated by CP-violation in the seesaw Yukawa interactions at the origin of the neutrino masses. These asymmetries are partly converted into a baryon asymmetry via the sphaleron processes. In this context, the question is whether these asymmetries could also be partially reconverted into a dark matter asymmetry, thereby dictating the dark matter relic density as well.

Several scenarios of this type have been proposed based on a higher-dimensional equilibration interaction~\cite{Kaplan:2009ag,Cohen:2009fz,Cui:2011qe,Ibe:2011hq,Servant:2013uwa,Boucenna:2015haa,Dhen:2015wra} or with renormalizable interactions~\cite{Dhen:2015wra}, see also \cite{Boucenna:2015haa}. All these scenarios have some concerns. An important experimental one comes from direct-detection experiments that have significantly improved their sensitivity in recent years~\cite{LZ:2024zvo}. These bounds preclude using Standard Model final states to induce fast dark matter annihilation processes that need to be more efficient than for WIMPs, so that the DM relic abundance is dominated by a particle-antiparticle asymmetry. For instance, a $\overline{DM} DM H^\dagger H$ Higgs portal interaction is fully excluded for perturbative interactions for the complex scalar singlet case \cite{EscuderoAbenza:2025cfj}. Other issues are theoretical. One is the need for a UV completion for scenarios based on higher-dimensional operators, which can be involved. Another concerns scenarios in which one assumes a $Z_2$ symmetry to stabilize the DM candidate and where the DM is hyperchargeless. In this case, a $DM^2$ quadratic term is allowed and would put any created DM asymmetry back to zero, unless very tiny.  Large charged lepton flavor violation can also arise generically in setups involving standard model leptons in the equilibration interaction, unless a large flavor hierarchy between lepton generations is assumed.

\section{The scenario: Minimal Baryogenesis induced Asymmetric Dark Matter} 
In this letter, we propose a new baryogenesis induced asymmetric dark matter scenario that is perfectly viable, is simple, and has signatures, see Fig.~\ref{fig:Mechanism_sketch} for a sketch. 
Our model is rather simple because it is based on the adjunction of only two new fields, and is based only on renormalizable interactions. This also maximizes its testability as all the physics at the origin of the dark matter asymmetry lies around the dark matter mass scale, or below. The two fields assumed are a complex singlet scalar $\phi$, and an inert Higgs doublet $H'$. 

In our scenario the dark matter today arises predominantly from an asymmetric population of $\phi$ particles. On the other hand, the new Higgs doublet  $H'=(H^+ \,,(H_0+iA_0)/\sqrt{2})^T$ is employed to 1) enable the transport of the SM asymmetries into the dark matter sector\footnote{\label{footimportant}In the presence of a baryon asymmetry and at $T>T_{\rm EW}$ all SM states but the $W^\pm$ have different numbers of particles and antiparticles including the Higgs doublet~\cite{Harvey:1990qw}.}, and 2) as a final state for $\phi \bar{\phi} \to H'\bar{H}' $ annihilations, in agreement with the lack of signals at direct detection experiments. 

 We assume a $Z_4$ DM stabilization symmetry, under which $\phi$ has charge $+1$ and $H'$ charge $-2$. This forbids a quadratic term in the DM field, $\phi^2$, which as already mentioned above, would diminish any asymmetry created, unless very tiny. Furthermore, this particle content and symmetry structure allows for two renormalizable particle-antiparticle asymmetry transfer channels: $\lambda_5 (H^\dagger H')^2$ and $\lambda_{\phi^2HH'}\phi^2H^\dagger H'$ (with $H$ being the SM Higgs doublet), whose interplay will lead to the new dynamical scenario we present here for generating the asymmetry. Note importantly that there are very few possible renormalizable interactions that could transfer a SM asymmetry into a DM asymmetry, and these two interactions are the simplest ones.

The $\lambda_5$ interaction that involves only one new BSM field cannot lead on its own to the observed relic density with DM being the lightest neutral component of the inert doublet $H'$. 
The lightest neutral component of a $H'$ inert doublet, $H_0$ or $A_0$, is excluded by direct detection experiments as a DM candidate, because it undergoes too fast $Z$-exchange scattering with nucleons, unless the latter is unavailable kinematically. As the $Z$ boson couples to $H_0A_0$, this will be the case provided that the neutral components split by $\delta m_{H_0} \equiv |m_{H_0}- m_{A_0}|\gtrsim 250\,{\rm keV} $~\cite{Bramante:2016rdh,Eby:2019mgs,Song:2021yar}. This will be accomplished after electroweak symmetry breaking (EWSB) if the $\lambda_5$ interaction is sizable enough:
\begin{equation}
    |\lambda_5|\gtrsim 1.65\cdot 10^{-6}\,\left(\frac{m_{H_0}}{200\,\hbox{GeV}}\right)\, \left(\frac{\delta m_{H_0}}{250\text{ keV}}\right)\,.
\label{eq:lambda5DDcondition}
\end{equation}
However, a $\lambda_5$ interaction satisfying this bound turns out to be still in thermal equilibrium when the $H'$ is already deeply non-relativistic, largely Boltzmann suppressing the $H'$ asymmetry, so that the resulting $H'$ asymmetry 
is too small to account for the DM observed relic density (explaining why we consider the singlet $\phi$ to be responsible for most of the relic density rather than $H'$), see \cite{Dhen:2015wra} and Eq.~(\ref{deltaH'}), below. 

The $\lambda_{\phi^2HH'}\phi^2H^\dagger H'$ interaction also cannot lead on its own to the observed DM relic density. Without $\lambda_5$, there is no mass splitting and $m_{H'}$ must be heavier than twice the mass of $\phi$ so that it is unstable. But in this case, as a result of the fact that without $\lambda_5$ the Lagrangian has a global $U(1)$ symmetry, even if a $\phi$ asymmetry is created, it will be erased by the $H'$ decay later on. However, the interplay of both $\lambda_{\phi^2HH'}$ and $\lambda_5$ interactions can easily do the job.

The relevant terms of the scalar potential are the following~\footnote{Note that the same scalar singlet-doublet content has been used in different contexts, including for asymmetric DM \cite{Dhen:2015wra} and symmetric DM~\cite{Belanger:2021lwd,Belanger:2022qxt,Cabral-Rosetti:2017mai}.}
\begin{align}\label{eq:potential}
    \!\!\!\!\!\!\!\! & V(H,H',\phi)= m_\phi^2 |\phi|^2  +m^2_{H'}  |H'|^2+\lambda_{\phi H} |\phi|^2  |H|^2 \\
    &+\lambda_{\phi H'} |\phi|^2 |H'|^2 
    +\lambda_{3} |H|^2 |H'|^2
    +\lambda_{4} |H^\dagger H'|^2\nonumber\\
    &+\left(\frac{\lambda_{\phi^2HH'}}{2} \phi^2  H^\dagger H' +h.c.\right)
    + \Big(\frac{\lambda_5}{2} (H^\dagger H')^2+h.c.\Big)\,. \nonumber
\end{align}
Note that $\lambda_{\phi^2HH'}'\phi^2  H H'^\dagger/2$ and $\lambda_{\phi4}(\phi^4+h.c.)/4!$ interactions are also allowed by the $Z_4$ symmetry. \modific{We have explicitly checked that unless both $|\lambda_{\phi^2HH'}'|$ and $|\lambda_{\phi^2HH'}|$ are equal within $\sim$10\%, or unless $|\lambda_{\phi 4}|$ is much bigger than the largest of these two, the produced dark matter asymmetry is not significantly reduced, see Supplemental Material~\ref{sec:appendices}~\cite{SuppMat}. For simplicity, in what follows, we neglect these other couplings.}\footnote{\label{foot_CP}Note that all interaction couplings in Eq.~\eqref{eq:potential} can be made real by an appropriate rephasing of the various fields, so that the only CP violation at the source of the DM asymmetry is the one at the origin of baryogenesis. Technically, only the secondary $\lambda_{\phi^2HH'}'$ and $\lambda_{\phi4}$ couplings can carry a physical phase but their effect will be irrelevant.}

There are several dynamical ways along which the DM asymmetry could be created a priori in this setup, depending mostly on the mass spectrum assumed and on which of the $\lambda_5$ and $\lambda_{\phi^2HH'}$ interactions reaches thermal equilibrium at some point. Several of them turn out to be excluded for various reasons, but one
regime turns out to work very well generically, which we present in this letter. It considers the  $m_{H'}<m_\phi $ mass spectrum and the case where the $\lambda_5$ interaction reaches thermal equilibrium,  whereas the $\lambda_{\phi^2HH'}$ never does, leading to an out-of-equilibrium production of the DM $\phi$ asymmetry. 

Before explaining how the produced $\phi$ asymmetry accounting for the observed DM relic density is created along this scenario, let us first discuss how a number of important constraints that appear to be fatal in many other potential setups are satisfied in this scenario.

a) \underline{Fate of the symmetric DM $\phi$ component}. To ensure that the DM symmetric component is sub-leading with respect to the asymmetric component turns out to be a very strong constraint for asymmetric dark matter scenarios. For instance the direct detection constraints exclude the possibility that the $\lambda_{\phi H}$ Higgs portal interaction successfully depletes the symmetric dark matter abundance~\cite{EscuderoAbenza:2025cfj}. However, if $m_{H'}<m_\phi$, the 
$\phi \bar{\phi}$ annihilation into $H'\bar{H}'$, from the $\lambda_{\phi H'}$ interaction
(which is less constrained) can do the job.

b) \underline{Fate of the symmetric DM $H'$ component}. If $m_{H'}<m_\phi$ as we assume,
DM is not only made of $\phi$ but also of the lightest neutral component of $H'$, since it is stable too. One should therefore wonder about the fate of its symmetric component. This will be subleading if  $m_{H'}< m_{H'}^{\rm EW}$, with $m_{H'}^{\rm EW}=580$~GeV, see~\cite{Bottaro:2022one,Cirelli:2005uq}, the value of $m_{H'}$ for which the electroweak symmetric annihilations of $H'$ leads to the observed dark matter abundance. Therefore we will consider that $m_{H'}< m_{H'}^{\rm EW}$.

c) \underline{Fate of the asymmetric DM $H'$ component}. If the $\lambda_5$ interaction is in thermal equilibrium, then $\mu_{H'}=\mu_H$, which implies that when this interaction enters thermal equilibrium, the following equality holds
\begin{equation}
    \frac{\Delta H'}{\Delta H}= \frac{Y_{H'}^{eq}}{Y_{H}^{eq}}  \propto e^{-m_{H'}/T}\,,
    \label{deltaH'}
\end{equation}
 where $Y_X$ refers to the number density, $n_X$, of a given species $X$, normalized to entropy, and $\Delta X \equiv Y_X-Y_{\bar{X}} $. For $T\gtrsim m_{H'}$ $\Delta H = \Delta H'$, but for $T\lesssim m_{H'}$ this gives instead a Boltzmann suppressed $H'$ asymmetry.

\begin{figure*}[!t]
\centering
 \hspace{-0.3cm} \includegraphics[width=0.33\textwidth]{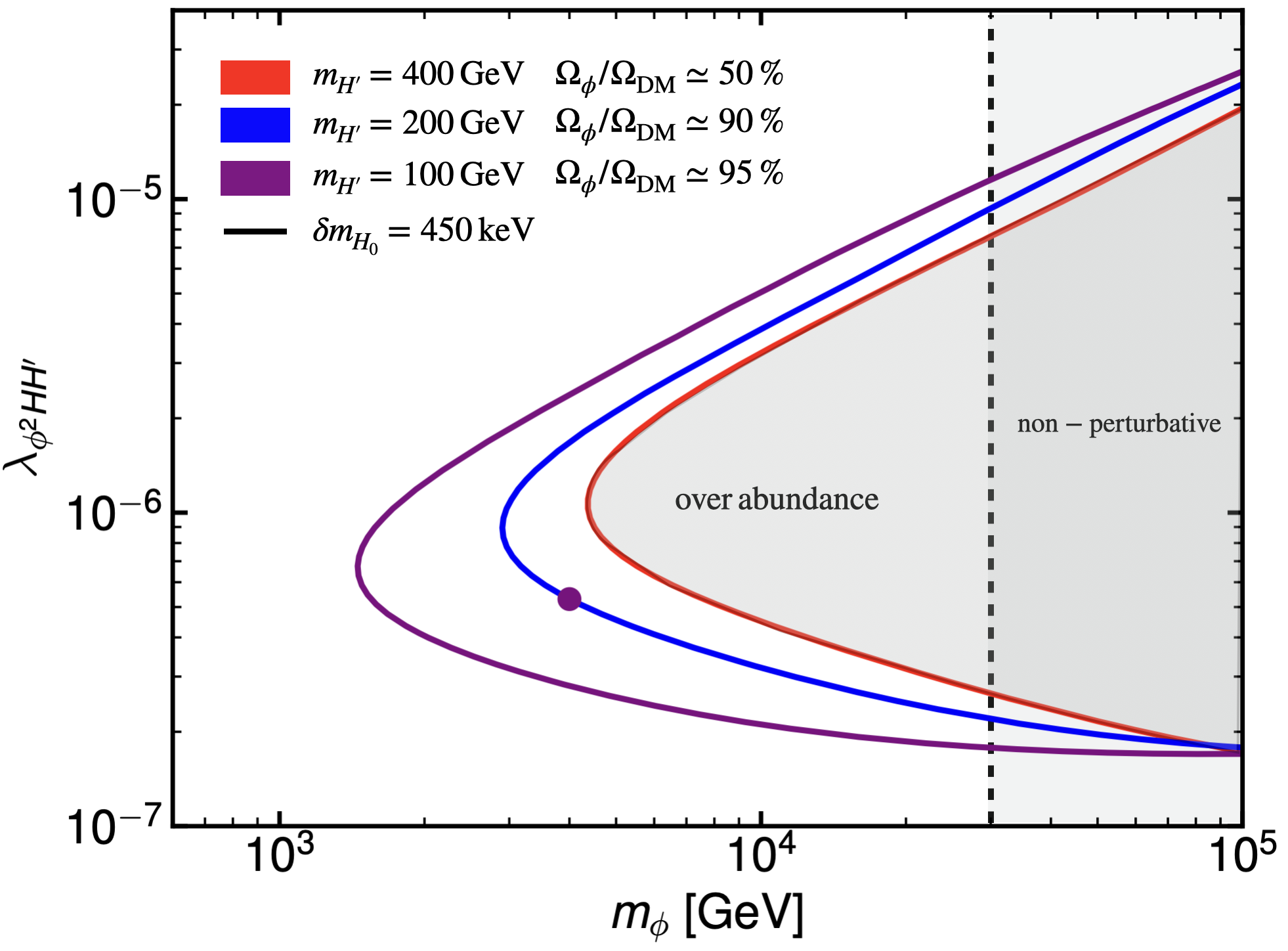} \hspace{-0.1cm} \includegraphics[width=0.33\textwidth]{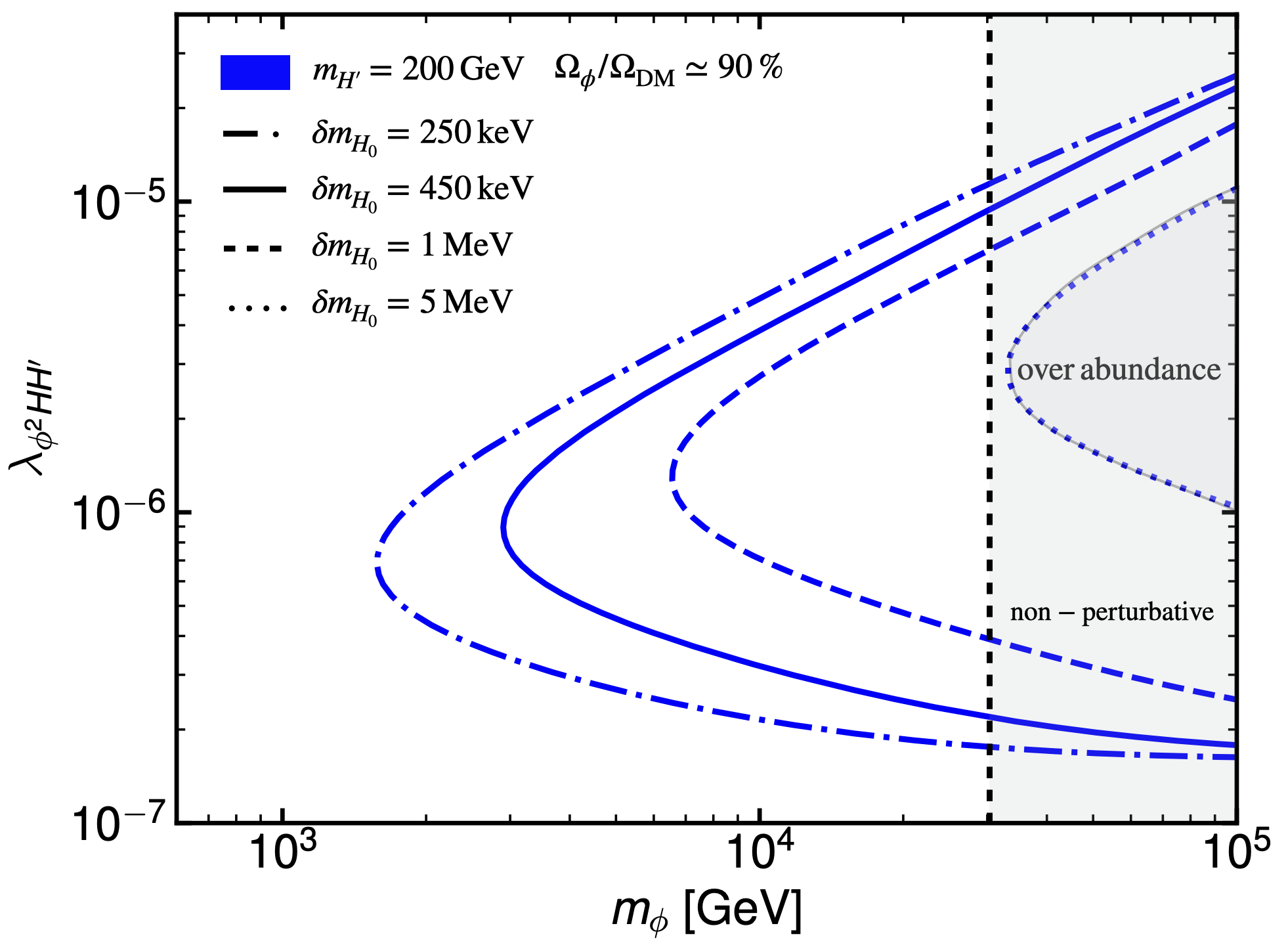} \hspace{-0.1cm}  \includegraphics[width=0.33\textwidth]{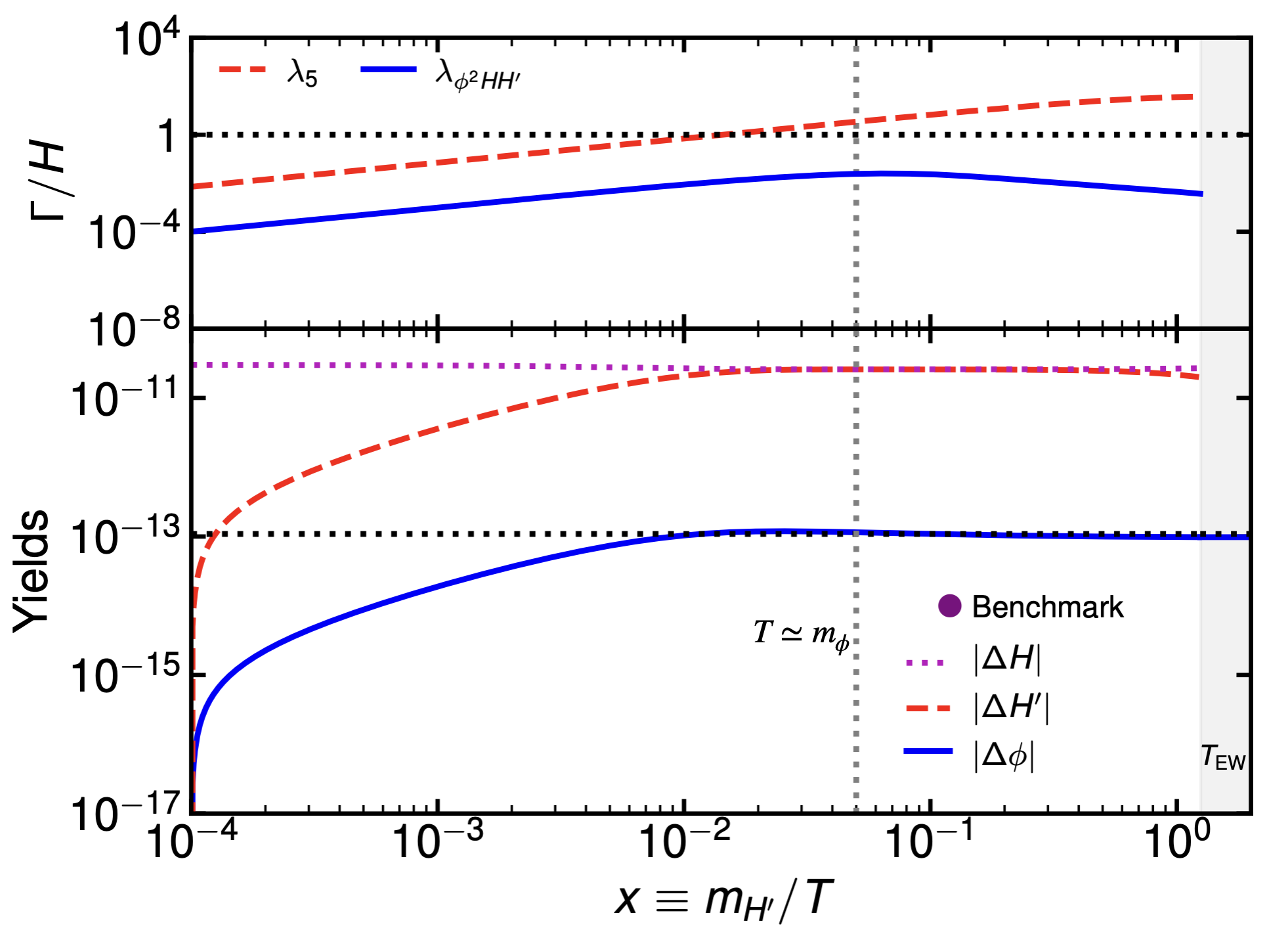}
\vspace{-0.2cm}
\caption{Parameter space in the $\lambda_{\phi^2HH'}$-$m_\phi$ plane, for various values of the doublet mass $m_{H'}$ (left) and of the neutral mass splitting $\delta m_{H_0} \equiv |m_{H_0}-m_{A_0}|$ (middle) that is directly related to the $\lambda_5$ coupling. For $m_\phi < 30\,{\rm TeV}$ the $\lambda_{\phi H'}$ coupling driving symmetric annihilations is expected to be perturbative~\cite{EscuderoAbenza:2025cfj}. In the right panel, we show the evolution for the benchmark point highlighted in the left panel as a purple dot. \modific{The upper sub-panel represents the two relevant rates and the lower one the relevant yields}. 
}
\label{fig:Mechanism_results}
\end{figure*}

The relation above holds approximately as long as chemical equilibrium is maintained which happens until relatively low temperatures, due to $H \bar{H}' \rightarrow \bar{H} H' $ scatterings which are not Boltzmann suppressed. In any case, the fate of the $H'$ asymmetry is to disappear as at the electroweak phase transition $T_{\rm EW}\simeq 160\,{\rm GeV}$ the $H'$ will split into two neutral and two charged states and the two neutral states will undergo very fast oscillations which will set the asymmetry to zero (see e.g.~\cite{Dhen:2015wra}).

d) \underline{Direct detection constraints on $\phi$ and $H'$}. Direct detection constraints on $\phi$ are trivially satisfied because $\phi$ interacts with the SM only through either the $\lambda_{\phi^2HH'}$ interaction (which we assume not to enter into thermal equilibrium and thus be small) or the Higgs portal interaction $\lambda_{\phi H}$ which is not assumed to play any role in the scenario and thus can be small. Similarly, $H'$ in our scenario also satisfies all direct detection constraints. As mentioned above we work in a regime where $H'\bar{H} \to H\bar{H}' $ thermalizes, which means that the $\lambda_5$ coupling satisfies Eq.~\eqref{eq:lambda5DDcondition}. Hence there is no dangerous tree-level $Z$ mediated interaction rate with nuclei. Only a very small loop-suppressed contribution survives, see e.g.~\cite{Bottaro:2022one}. Importantly, $\lambda_3$ and $\lambda_4$ lead to direct detection via Higgs exchange as well, but are not needed to ensure a fast enough $H'$ annihilation, as below 580 GeV this is ensured by the electroweak $H'\bar{H}'\to W^+W^-$ channel.

While bounds are satisfied, the scenario perfectly allows that the values of these three couplings are below but close to the present direct detection sensitivity, so that $\phi$ or $H'$ particles may be observed in the near future. The relevant interactions with nuclei are spin independent (see e.g.~\cite{EscuderoAbenza:2025cfj}) and recasting the latest limit from LZ~\cite{LZ:2024zvo} one finds:
\begin{align}
    |\lambda_{S h}|< 0.057 \,\left(\frac{m_S}{\rm TeV}\right)^{3/2}\,\sqrt{\frac{1}{f_S}}\,,\quad \text{[LZ limit]}
\end{align}
where $S = \phi,\,H_0/A_0$, with $f_S$ being the fraction of the dark matter energy density in the form of $S$. For both dark matter components we have: $\lambda_{\phi h} = \lambda_{\phi H}$, $  \lambda_{H_0 h} = \lambda_3+\lambda_4+\lambda_5 \simeq \lambda_3 - \frac{2}{v_H^2}(m_{H^\pm}^2-m_{H_0}^2)$, $
    \lambda_{A_0 h} = \lambda_3+\lambda_4-\lambda_5 =\lambda_3 - \frac{2}{v_H^2}(m_{H^\pm}^2-m_{A_0}^2)$, with $v_H = 246\,{\rm GeV}$.

Interestingly, assuming no strong cancellation between $\lambda_3$ and $\lambda_4$ we can impose a limit for each of these couplings, and the one for $\lambda_4$ actually leads to a limit on the induced mass splitting between the charged and neutral components, as follows:
\begin{align}
    m_{H^\pm}-m_{H_0}\simeq -\frac{\lambda_4 v_H^2}{4m_{H'}} \lesssim 1\,{\rm GeV}\,.\quad \text{[LZ},\, \lambda_3=0 \text{]}
\end{align}
\modific{As we will see later, this mass splitting dictates the lifetime of $H^\pm$ at colliders and hence leads to an interesting correlation between two different signatures to be measured by two distinct experiments.}

\section{Asymmetry Generation and Parameter Space}


At a temperature much larger than all masses involved, none of the \modific{asymmetric portal} interactions are in thermal equilibrium as their rate scales as $\Gamma \sim \lambda^2 T$. As a result, a $\phi$ asymmetry, $\Delta \phi$, is created out of equilibrium by the processes induced by the $\lambda_{\phi^2HH'}$ interaction, $\phi \bar{H}  \rightarrow \bar{\phi} \bar{H}'$ and $H \bar{H}' \rightarrow \phi\phi$ (and conjugate processes). Considering the system in the ultrarelativistic limit, the $\phi$ asymmetry evolves as: 
\begin{align}\label{eq:simple_phi_equation}
\frac{d\Delta \phi}{dx} \simeq \,\, &(\Delta H - \Delta H'-4\Delta \phi) \,\times \nonumber \\
   & 3\frac{|\lambda_{\phi^2HH'}|^2}{128\pi} \frac{{m_{\rm Pl}}}{1.66\pi^2 \sqrt{g_\star} m_{H'}} \,,
\end{align}
where $x \equiv m_{H'}/T$, $m_{\rm Pl} = 1.22\times 10^{19}\,{\rm GeV}$, and $g_\star \simeq 106$. 
At early times, $\Delta H' = \Delta \phi = 0$ and the $\Delta \phi$ asymmetry is generated through freeze-in from the Higgs $\Delta H$ asymmetry generated via baryogenesis.

\modific{The parameter space of the model in the $\lambda_{\phi^2HH'}$-$m_\phi$ plane is shown in Fig.~\ref{fig:Mechanism_results}. } From this figure one observes that the right dark matter abundance is obtained in two branches for $\lambda_{\phi^2 HH'}$. The reason is simple: the lower branch corresponds to the case where the $\Delta \phi$ asymmetry is generated via almost freeze-in $\lambda_{\phi^2 HH'}$ coupling alone while the $\lambda_5$ interactions are still out of equilibrium. \modific{This can be explicitly understood from the right panel which shows the temperature evolution of the asymmetries and rates for a successful set of couplings and masses.} The upper branch corresponds on the other hand to a case where $\lambda_{\phi^2 HH'}$ interactions are out-of-equilibrium but not too far from the Hubble rate at a time where $\lambda_5$ interactions are deep in thermal equilibrium and $\Delta H' \simeq \Delta H$. This in turn actually suppresses the $\Delta \phi$ asymmetry and can lead to the correct dark matter abundance.

\vspace{0.1cm}

Note also that reducing $m_\phi$, and keeping all other parameters fixed, leads to a smaller asymmetry, because in this case, both the $\lambda_5$ and $\lambda_{\phi^2HH'}$ interaction thermalize more at $T\sim m_{\phi}$, whereas for too large a value of $m_\phi$  the $\phi$ symmetric component cannot be suppressed enough with perturbative couplings. The efficiency of the mechanism then requires $m_\phi \gtrsim 400\,{\rm GeV}$ and also a mild hierarchy between the $\phi$ and $H'$ masses, $m_{\phi} \gtrsim 10\,m_{H'}$. This last point is clearly illustrated in the left panel of Fig.~\ref{fig:Mechanism_results} and we refer to the Supplemental Material~\ref{sec:appendices} for additional explanations and benchmark points illustrating all these features and which includes Refs.~\cite{Edsjo:1997bg,Planck:2018vyg}.

\section{Signatures}

\noindent All in all  
\modific{the parameter space of the scenario is}:
\begin{subequations}
\begin{align}
 [\rm eff.]\,\, \,   400\,\hbox{GeV}&\lesssim \,\,\,\,m_\phi\,\,\,\,\lesssim 30~\hbox{TeV}\,,\, \,\,\,\,\,[\rm pert.]\\[4pt]
 [{\rm LEP}] \,\,\,\,\,\,  46\, \hbox{GeV}&\lesssim \,\,\,m_{H'}\,\,<580~\hbox{GeV} \,,\,\, [{\rm DMab}]\\[4pt]
    m_\phi &\gtrsim 10 \,m_{H'}\,,\,\,\,\,[\rm eff.]\\[4pt]
  [\rm DD] \,\,\,\,\,   250\,\hbox{keV}&\lesssim \,\,\delta m_{H_0}\lesssim 5~\hbox{MeV}\,, \,\,\,\,\,\,\,[\rm eff.]\label{eq:deltaH0}\\[4pt] 
 m_{H^\pm}-m_{H_0} &\lesssim \modific{1\,{\rm GeV}\,,\quad \text{[DD},\, ({\rm for}\,\lambda_3=0) \text{]}}\label{eq:deltaHp}\\[4pt]
\lambda_{\phi H'}  \,>0.35\,& m_\phi/(700\,{\rm GeV})\,.\,\,[\bar{\phi} \phi \,{\rm annihilation}]
\end{align}
\end{subequations} where the origin of each bound has been indicated. \modific{The lower bound on $m_{H'}$ comes from the invisible decay of the $Z$~\cite{Belyaev:2016lok} and quite remarkably a charged particle in line with Eqs.~\eqref{eq:deltaH0} and \eqref{eq:deltaHp} evades all LEP searches for masses above 46 GeV,  see~\cite{Belyaev:2016lok,Blinov:2015qva}}. The other bounds are described in previous sections (eff. refers to the efficiency of the mechanism, see Fig.~\ref{fig:Mechanism_results}). Note that $\Omega_{H_0}h^2 = 0.12\,(m_{H_0}/580\,{\rm GeV})^2$, hence the lighter is $H'$, the more asymmetric is dark matter and also that the efficiency of the mechanism requires a mild mass hierarchy between the two states in the dark matter sector, $m_{\phi} \gtrsim 10\,m_{H'}$.  \\

A key prediction of the scenario is that the second inert Higgs doublet is light and therefore has clear discovery possibilities at the LHC, HL-LHC, and future colliders. In particular, due to the compressed particle spectra in our scenario, $H'$ can lead to disappearing \modific{and displaced} charged tracks, $H^+\to A_0 + \pi^+$. \modific{These types of signatures represent key channels of the flagship searches at the LHC and very recently ATLAS and CMS have been able to supersede the LEP sensitivity for degenerate Higgsinos, see~\cite{ATLAS:2022rme,ATLAS:2024umc,ATLAS:2025lhc,CMS:2025qkk,CMS:2025mie}. The inert doublet being a scalar has a smaller production cross section by a factor of $\sim 8$~\cite{Belyaev:2020wok} and represents the next natural target. Critically, for a general inert doublet model, a light inert doublet would only be a sub-leading dark matter component, whereas in our case it plays a major role as it is the portal to dark matter asymmetry. Furthermore, its lifetime is strongly correlated with the charged-neutral mass splitting and in the absence of strong cancellations it is directly controlled by the scattering with nuclei of $H_0/A_0$. Given that $H'$ may be as light as $46\,{\rm GeV}$ and hence lighter than a xenon nucleus, it is possible that the next generation of direct detection experiments~\cite{XLZD:2024nsu,PANDA-X:2024dlo} could identify both its mass and its scattering rate and hence make a strong connection with the signatures at the HL-LHC. A detailed study of this phenomenology is beyond the scope of the present work, but depending upon $m_{H^\pm}-m_{H_0}$ the LHC may be already sensitive up to $m_{H'}\sim 170\,{\rm GeV}$~\cite{Belyaev:2020wok}, or up to $250\,{\rm GeV}$ by the end of the HL-LHC~\cite{ATLAS:2022rme,CMS:2023mny}. Since the displaced and disappearing charged track signatures are rather unique, it is quite plausible that new and bold trigger strategies could probe the parameter space even further~\cite{deBlas:2025gyz}.}

\section{Conclusions and Discussions}
In this Letter, we have investigated whether the DM relic density can originate from a primordial asymmetry directly related to the baryogenesis induced SM asymmetries, and showed that—despite strong constraints—this can be achieved in a simple, renormalizable, viable setup with a scalar singlet and a second inert scalar doublet. Along this scenario the asymmetry is slowly produced in an out-of-equilibrium way, through an ``asymmetric freeze-in''. 
The minimality of the scenario and the fact that it is based only on renormalizable interactions make this possibility rather unique. It leads to several possibilities of tests in the near future, including in particular \modific{the existence of a rather light inert scalar doublet with mass $46\,{\rm GeV} \lesssim m_{H'}\lesssim 580\,{\rm GeV}$, which can lead to rather unique signatures at the HL-LHC}, and possibilities for DM direct detection through a Higgs portal interaction for both dark matter components. Also, as the freeze-in production naturally converts only a small fraction of the lepton/baryon asymmetry into a DM asymmetry, the DM mass scale is well above the usual naive (and often experimentally problematic)  $m_{\rm DM} \sim 5\times m_p$  expectation which holds in many asymmetric dark matter scenarios from the observed ratio $\Omega_{\rm DM} /\Omega_{b} \simeq 5.3$. 

Symmetric dark matter freeze-in, unlike symmetric freeze-out, suffers from an initial condition dependence, as it is based on the assumption that there is no DM population with which to start. For asymmetric freeze-in instead, one assumes only that there is no DM asymmetry with which to start, which is a much weaker assumption, in the same way as for baryogenesis where one assumes that there is no baryon asymmetry initially (at the end of inflation)\footnote{See footnote~\ref{foot_CP}.}. In practice this makes a crucial theoretical and experimental difference: along the asymmetric freeze-in scenario the DM particles are perfectly allowed to couple efficiently to SM states (for example through the various quartic couplings), whereas in the symmetric freeze-in scenario all the portal interactions have to be assumed to be very tiny.  Related to this discussion, the scenario we consider \modific{features a hierarchy between the symmetric couplings and the asymmetric ones. That is, while $\lambda_5, \,\lambda_{\phi^2HH'}$ need to be $\sim 10^{-6}$, other quartics can be fairly large: $\lambda_{\phi H'} > 0.35\,m_{\phi}/(700\,\,{\rm GeV})$, $\lambda_3,\lambda_4\lesssim 0.01$, $ \lambda_{\phi H} \lesssim 0.06 \,(m_\phi/{\rm TeV})^{3/2}$. Even though small asymmetric couplings are technically natural, it would be worth investigating if this intriguing hierarchy could find an appealing theoretical explanation.}



\begin{acknowledgments}
We are grateful to Tim Cohen for pointing out that our quartic portal couplings are technically natural and to our referees for questioning the possible hierarchies in our scalar potential. The work of TH is supported by the Belgian IISN convention 4.4503.15 and by the Brussels Laboratory of the
Universe - BLU-ULB. C.~H. is funded by the Generalitat Valenciana under Plan Gen-T via CDEIGENT grant No. CIDEIG/2022/16.  C.~H. also acknowledges partial support from the Belgian IISN convention 4.4503.15 as well as the Spanish grants PID2023-147306NB-I00 and CEX2023-001292-S (MCIU/AEI/10.13039/501100011033).
\end{acknowledgments}

\bibliography{biblio}

\begin{thebibliography}{41}%
\makeatletter
\providecommand \@ifxundefined [1]{%
 \@ifx{#1\undefined}
}%
\providecommand \@ifnum [1]{%
 \ifnum #1\expandafter \@firstoftwo
 \else \expandafter \@secondoftwo
 \fi
}%
\providecommand \@ifx [1]{%
 \ifx #1\expandafter \@firstoftwo
 \else \expandafter \@secondoftwo
 \fi
}%
\providecommand \natexlab [1]{#1}%
\providecommand \enquote  [1]{``#1''}%
\providecommand \bibnamefont  [1]{#1}%
\providecommand \bibfnamefont [1]{#1}%
\providecommand \citenamefont [1]{#1}%
\providecommand \href@noop [0]{\@secondoftwo}%
\providecommand \href [0]{\begingroup \@sanitize@url \@href}%
\providecommand \@href[1]{\@@startlink{#1}\@@href}%
\providecommand \@@href[1]{\endgroup#1\@@endlink}%
\providecommand \@sanitize@url [0]{\catcode `\\12\catcode `\$12\catcode
  `\&12\catcode `\#12\catcode `\^12\catcode `\_12\catcode `\%12\relax}%
\providecommand \@@startlink[1]{}%
\providecommand \@@endlink[0]{}%
\providecommand \url  [0]{\begingroup\@sanitize@url \@url }%
\providecommand \@url [1]{\endgroup\@href {#1}{\urlprefix }}%
\providecommand \urlprefix  [0]{URL }%
\providecommand \Eprint [0]{\href }%
\providecommand \doibase [0]{http://dx.doi.org/}%
\providecommand \selectlanguage [0]{\@gobble}%
\providecommand \bibinfo  [0]{\@secondoftwo}%
\providecommand \bibfield  [0]{\@secondoftwo}%
\providecommand \translation [1]{[#1]}%
\providecommand \BibitemOpen [0]{}%
\providecommand \bibitemStop [0]{}%
\providecommand \bibitemNoStop [0]{.\EOS\space}%
\providecommand \EOS [0]{\spacefactor3000\relax}%
\providecommand \BibitemShut  [1]{\csname bibitem#1\endcsname}%
\let\auto@bib@innerbib\@empty
\bibitem [{\citenamefont {Kolb}\ and\ \citenamefont
  {Turner}(1990)}]{Kolb:1990vq}%
  \BibitemOpen
  \bibfield  {author} {\bibinfo {author} {\bibfnamefont {E.~W.}\ \bibnamefont
  {Kolb}}\ and\ \bibinfo {author} {\bibfnamefont {M.~S.}\ \bibnamefont
  {Turner}},\ }\href {\doibase 10.1201/9780429492860} {\emph {\bibinfo {title}
  {{The Early Universe}}}},\ Vol.~\bibinfo {volume} {69}\ (\bibinfo {year}
  {1990})\BibitemShut {NoStop}%
\bibitem [{\citenamefont {Rubakov}\ and\ \citenamefont
  {Gorbunov}(2017)}]{Rubakov:2017xzr}%
  \BibitemOpen
  \bibfield  {author} {\bibinfo {author} {\bibfnamefont {V.~A.}\ \bibnamefont
  {Rubakov}}\ and\ \bibinfo {author} {\bibfnamefont {D.~S.}\ \bibnamefont
  {Gorbunov}},\ }\href {\doibase 10.1142/10447} {\emph {\bibinfo {title}
  {{Introduction to the Theory of the Early Universe}: {Hot big bang
  theory}}}}\ (\bibinfo  {publisher} {World Scientific},\ \bibinfo {address}
  {Singapore},\ \bibinfo {year} {2017})\BibitemShut {NoStop}%
\bibitem [{\citenamefont {Cirelli}\ \emph {et~al.}(2024)\citenamefont
  {Cirelli}, \citenamefont {Strumia},\ and\ \citenamefont
  {Zupan}}]{Cirelli:2024ssz}%
  \BibitemOpen
  \bibfield  {author} {\bibinfo {author} {\bibfnamefont {M.}~\bibnamefont
  {Cirelli}}, \bibinfo {author} {\bibfnamefont {A.}~\bibnamefont {Strumia}}, \
  and\ \bibinfo {author} {\bibfnamefont {J.}~\bibnamefont {Zupan}},\
  }\href@noop {} {\  (\bibinfo {year} {2024})},\ \Eprint
  {http://arxiv.org/abs/2406.01705} {arXiv:2406.01705 [hep-ph]} \BibitemShut
  {NoStop}%
\bibitem [{\citenamefont {Zurek}(2014)}]{Zurek:2013wia}%
  \BibitemOpen
  \bibfield  {author} {\bibinfo {author} {\bibfnamefont {K.~M.}\ \bibnamefont
  {Zurek}},\ }\href {\doibase 10.1016/j.physrep.2013.12.001} {\bibfield
  {journal} {\bibinfo  {journal} {Phys. Rept.}\ }\textbf {\bibinfo {volume}
  {537}},\ \bibinfo {pages} {91} (\bibinfo {year} {2014})},\ \Eprint
  {http://arxiv.org/abs/1308.0338} {arXiv:1308.0338 [hep-ph]} \BibitemShut
  {NoStop}%
\bibitem [{\citenamefont {Petraki}\ and\ \citenamefont
  {Volkas}(2013)}]{Petraki:2013wwa}%
  \BibitemOpen
  \bibfield  {author} {\bibinfo {author} {\bibfnamefont {K.}~\bibnamefont
  {Petraki}}\ and\ \bibinfo {author} {\bibfnamefont {R.~R.}\ \bibnamefont
  {Volkas}},\ }\href {\doibase 10.1142/S0217751X13300287} {\bibfield  {journal}
  {\bibinfo  {journal} {Int. J. Mod. Phys. A}\ }\textbf {\bibinfo {volume}
  {28}},\ \bibinfo {pages} {1330028} (\bibinfo {year} {2013})},\ \Eprint
  {http://arxiv.org/abs/1305.4939} {arXiv:1305.4939 [hep-ph]} \BibitemShut
  {NoStop}%
\bibitem [{\citenamefont {Boucenna}\ and\ \citenamefont
  {Morisi}(2014)}]{Boucenna:2013wba}%
  \BibitemOpen
  \bibfield  {author} {\bibinfo {author} {\bibfnamefont {S.~M.}\ \bibnamefont
  {Boucenna}}\ and\ \bibinfo {author} {\bibfnamefont {S.}~\bibnamefont
  {Morisi}},\ }\href {\doibase 10.3389/fphy.2013.00033} {\bibfield  {journal}
  {\bibinfo  {journal} {Front. in Phys.}\ }\textbf {\bibinfo {volume} {1}},\
  \bibinfo {pages} {33} (\bibinfo {year} {2014})},\ \Eprint
  {http://arxiv.org/abs/1310.1904} {arXiv:1310.1904 [hep-ph]} \BibitemShut
  {NoStop}%
\bibitem [{\citenamefont {Davoudiasl}\ and\ \citenamefont
  {Mohapatra}(2012)}]{Davoudiasl:2012uw}%
  \BibitemOpen
  \bibfield  {author} {\bibinfo {author} {\bibfnamefont {H.}~\bibnamefont
  {Davoudiasl}}\ and\ \bibinfo {author} {\bibfnamefont {R.~N.}\ \bibnamefont
  {Mohapatra}},\ }\href {\doibase 10.1088/1367-2630/14/9/095011} {\bibfield
  {journal} {\bibinfo  {journal} {New J. Phys.}\ }\textbf {\bibinfo {volume}
  {14}},\ \bibinfo {pages} {095011} (\bibinfo {year} {2012})},\ \Eprint
  {http://arxiv.org/abs/1203.1247} {arXiv:1203.1247 [hep-ph]} \BibitemShut
  {NoStop}%
\bibitem [{\citenamefont {Fukugita}\ and\ \citenamefont
  {Yanagida}(1986)}]{Fukugita:1986hr}%
  \BibitemOpen
  \bibfield  {author} {\bibinfo {author} {\bibfnamefont {M.}~\bibnamefont
  {Fukugita}}\ and\ \bibinfo {author} {\bibfnamefont {T.}~\bibnamefont
  {Yanagida}},\ }\href {\doibase 10.1016/0370-2693(86)91126-3} {\bibfield
  {journal} {\bibinfo  {journal} {Phys. Lett. B}\ }\textbf {\bibinfo {volume}
  {174}},\ \bibinfo {pages} {45} (\bibinfo {year} {1986})}\BibitemShut
  {NoStop}%
\bibitem [{\citenamefont {Kaplan}\ \emph {et~al.}(2009)\citenamefont {Kaplan},
  \citenamefont {Luty},\ and\ \citenamefont {Zurek}}]{Kaplan:2009ag}%
  \BibitemOpen
  \bibfield  {author} {\bibinfo {author} {\bibfnamefont {D.~E.}\ \bibnamefont
  {Kaplan}}, \bibinfo {author} {\bibfnamefont {M.~A.}\ \bibnamefont {Luty}}, \
  and\ \bibinfo {author} {\bibfnamefont {K.~M.}\ \bibnamefont {Zurek}},\ }\href
  {\doibase 10.1103/PhysRevD.79.115016} {\bibfield  {journal} {\bibinfo
  {journal} {Phys. Rev. D}\ }\textbf {\bibinfo {volume} {79}},\ \bibinfo
  {pages} {115016} (\bibinfo {year} {2009})},\ \Eprint
  {http://arxiv.org/abs/0901.4117} {arXiv:0901.4117 [hep-ph]} \BibitemShut
  {NoStop}%
\bibitem [{\citenamefont {Cohen}\ and\ \citenamefont
  {Zurek}(2010)}]{Cohen:2009fz}%
  \BibitemOpen
  \bibfield  {author} {\bibinfo {author} {\bibfnamefont {T.}~\bibnamefont
  {Cohen}}\ and\ \bibinfo {author} {\bibfnamefont {K.~M.}\ \bibnamefont
  {Zurek}},\ }\href {\doibase 10.1103/PhysRevLett.104.101301} {\bibfield
  {journal} {\bibinfo  {journal} {Phys. Rev. Lett.}\ }\textbf {\bibinfo
  {volume} {104}},\ \bibinfo {pages} {101301} (\bibinfo {year} {2010})},\
  \Eprint {http://arxiv.org/abs/0909.2035} {arXiv:0909.2035 [hep-ph]}
  \BibitemShut {NoStop}%
\bibitem [{\citenamefont {Cui}\ \emph {et~al.}(2011)\citenamefont {Cui},
  \citenamefont {Randall},\ and\ \citenamefont {Shuve}}]{Cui:2011qe}%
  \BibitemOpen
  \bibfield  {author} {\bibinfo {author} {\bibfnamefont {Y.}~\bibnamefont
  {Cui}}, \bibinfo {author} {\bibfnamefont {L.}~\bibnamefont {Randall}}, \ and\
  \bibinfo {author} {\bibfnamefont {B.}~\bibnamefont {Shuve}},\ }\href
  {\doibase 10.1007/JHEP08(2011)073} {\bibfield  {journal} {\bibinfo  {journal}
  {JHEP}\ }\textbf {\bibinfo {volume} {08}},\ \bibinfo {pages} {073} (\bibinfo
  {year} {2011})},\ \Eprint {http://arxiv.org/abs/1106.4834} {arXiv:1106.4834
  [hep-ph]} \BibitemShut {NoStop}%
\bibitem [{\citenamefont {Ibe}\ \emph {et~al.}(2012)\citenamefont {Ibe},
  \citenamefont {Matsumoto},\ and\ \citenamefont {Yanagida}}]{Ibe:2011hq}%
  \BibitemOpen
  \bibfield  {author} {\bibinfo {author} {\bibfnamefont {M.}~\bibnamefont
  {Ibe}}, \bibinfo {author} {\bibfnamefont {S.}~\bibnamefont {Matsumoto}}, \
  and\ \bibinfo {author} {\bibfnamefont {T.~T.}\ \bibnamefont {Yanagida}},\
  }\href {\doibase 10.1016/j.physletb.2012.01.032} {\bibfield  {journal}
  {\bibinfo  {journal} {Phys. Lett. B}\ }\textbf {\bibinfo {volume} {708}},\
  \bibinfo {pages} {112} (\bibinfo {year} {2012})},\ \Eprint
  {http://arxiv.org/abs/1110.5452} {arXiv:1110.5452 [hep-ph]} \BibitemShut
  {NoStop}%
\bibitem [{\citenamefont {Servant}\ and\ \citenamefont
  {Tulin}(2013)}]{Servant:2013uwa}%
  \BibitemOpen
  \bibfield  {author} {\bibinfo {author} {\bibfnamefont {G.}~\bibnamefont
  {Servant}}\ and\ \bibinfo {author} {\bibfnamefont {S.}~\bibnamefont
  {Tulin}},\ }\href {\doibase 10.1103/PhysRevLett.111.151601} {\bibfield
  {journal} {\bibinfo  {journal} {Phys. Rev. Lett.}\ }\textbf {\bibinfo
  {volume} {111}},\ \bibinfo {pages} {151601} (\bibinfo {year} {2013})},\
  \Eprint {http://arxiv.org/abs/1304.3464} {arXiv:1304.3464 [hep-ph]}
  \BibitemShut {NoStop}%
\bibitem [{\citenamefont {Boucenna}\ \emph {et~al.}(2015)\citenamefont
  {Boucenna}, \citenamefont {Krauss},\ and\ \citenamefont
  {Nardi}}]{Boucenna:2015haa}%
  \BibitemOpen
  \bibfield  {author} {\bibinfo {author} {\bibfnamefont {S.~M.}\ \bibnamefont
  {Boucenna}}, \bibinfo {author} {\bibfnamefont {M.~B.}\ \bibnamefont
  {Krauss}}, \ and\ \bibinfo {author} {\bibfnamefont {E.}~\bibnamefont
  {Nardi}},\ }\href {\doibase 10.1016/j.physletb.2015.06.080} {\bibfield
  {journal} {\bibinfo  {journal} {Phys. Lett. B}\ }\textbf {\bibinfo {volume}
  {748}},\ \bibinfo {pages} {191} (\bibinfo {year} {2015})},\ \Eprint
  {http://arxiv.org/abs/1503.01119} {arXiv:1503.01119 [hep-ph]} \BibitemShut
  {NoStop}%
\bibitem [{\citenamefont {Dhen}\ and\ \citenamefont
  {Hambye}(2015)}]{Dhen:2015wra}%
  \BibitemOpen
  \bibfield  {author} {\bibinfo {author} {\bibfnamefont {M.}~\bibnamefont
  {Dhen}}\ and\ \bibinfo {author} {\bibfnamefont {T.}~\bibnamefont {Hambye}},\
  }\href {\doibase 10.1103/PhysRevD.92.075013} {\bibfield  {journal} {\bibinfo
  {journal} {Phys. Rev. D}\ }\textbf {\bibinfo {volume} {92}},\ \bibinfo
  {pages} {075013} (\bibinfo {year} {2015})},\ \Eprint
  {http://arxiv.org/abs/1503.03444} {arXiv:1503.03444 [hep-ph]} \BibitemShut
  {NoStop}%
\bibitem [{\citenamefont {Aalbers}\ \emph {et~al.}(2025)\citenamefont {Aalbers}
  \emph {et~al.}}]{LZ:2024zvo}%
  \BibitemOpen
  \bibfield  {author} {\bibinfo {author} {\bibfnamefont {J.}~\bibnamefont
  {Aalbers}} \emph {et~al.} (\bibinfo {collaboration} {LZ}),\ }\href {\doibase
  10.1103/4dyc-z8zf} {\bibfield  {journal} {\bibinfo  {journal} {Phys. Rev.
  Lett.}\ }\textbf {\bibinfo {volume} {135}},\ \bibinfo {pages} {011802}
  (\bibinfo {year} {2025})},\ \Eprint {http://arxiv.org/abs/2410.17036}
  {arXiv:2410.17036 [hep-ex]} \BibitemShut {NoStop}%
\bibitem [{\citenamefont {Escudero~Abenza}\ and\ \citenamefont
  {Hambye}(2025)}]{EscuderoAbenza:2025cfj}%
  \BibitemOpen
  \bibfield  {author} {\bibinfo {author} {\bibfnamefont {M.}~\bibnamefont
  {Escudero~Abenza}}\ and\ \bibinfo {author} {\bibfnamefont {T.}~\bibnamefont
  {Hambye}},\ }\href {\doibase 10.1016/j.physletb.2025.139696} {\bibfield
  {journal} {\bibinfo  {journal} {Phys. Lett. B}\ }\textbf {\bibinfo {volume}
  {868}},\ \bibinfo {pages} {139696} (\bibinfo {year} {2025})},\ \Eprint
  {http://arxiv.org/abs/2505.02408} {arXiv:2505.02408 [hep-ph]} \BibitemShut
  {NoStop}%
\bibitem [{\citenamefont {Harvey}\ and\ \citenamefont
  {Turner}(1990)}]{Harvey:1990qw}%
  \BibitemOpen
  \bibfield  {author} {\bibinfo {author} {\bibfnamefont {J.~A.}\ \bibnamefont
  {Harvey}}\ and\ \bibinfo {author} {\bibfnamefont {M.~S.}\ \bibnamefont
  {Turner}},\ }\href {\doibase 10.1103/PhysRevD.42.3344} {\bibfield  {journal}
  {\bibinfo  {journal} {Phys. Rev. D}\ }\textbf {\bibinfo {volume} {42}},\
  \bibinfo {pages} {3344} (\bibinfo {year} {1990})}\BibitemShut {NoStop}%
\bibitem [{\citenamefont {Bramante}\ \emph {et~al.}(2016)\citenamefont
  {Bramante}, \citenamefont {Fox}, \citenamefont {Kribs},\ and\ \citenamefont
  {Martin}}]{Bramante:2016rdh}%
  \BibitemOpen
  \bibfield  {author} {\bibinfo {author} {\bibfnamefont {J.}~\bibnamefont
  {Bramante}}, \bibinfo {author} {\bibfnamefont {P.~J.}\ \bibnamefont {Fox}},
  \bibinfo {author} {\bibfnamefont {G.~D.}\ \bibnamefont {Kribs}}, \ and\
  \bibinfo {author} {\bibfnamefont {A.}~\bibnamefont {Martin}},\ }\href
  {\doibase 10.1103/PhysRevD.94.115026} {\bibfield  {journal} {\bibinfo
  {journal} {Phys. Rev. D}\ }\textbf {\bibinfo {volume} {94}},\ \bibinfo
  {pages} {115026} (\bibinfo {year} {2016})},\ \Eprint
  {http://arxiv.org/abs/1608.02662} {arXiv:1608.02662 [hep-ph]} \BibitemShut
  {NoStop}%
\bibitem [{\citenamefont {Eby}\ \emph {et~al.}(2019)\citenamefont {Eby},
  \citenamefont {Fox}, \citenamefont {Harnik},\ and\ \citenamefont
  {Kribs}}]{Eby:2019mgs}%
  \BibitemOpen
  \bibfield  {author} {\bibinfo {author} {\bibfnamefont {J.}~\bibnamefont
  {Eby}}, \bibinfo {author} {\bibfnamefont {P.~J.}\ \bibnamefont {Fox}},
  \bibinfo {author} {\bibfnamefont {R.}~\bibnamefont {Harnik}}, \ and\ \bibinfo
  {author} {\bibfnamefont {G.~D.}\ \bibnamefont {Kribs}},\ }\href {\doibase
  10.1007/JHEP09(2019)115} {\bibfield  {journal} {\bibinfo  {journal} {JHEP}\
  }\textbf {\bibinfo {volume} {09}},\ \bibinfo {pages} {115} (\bibinfo {year}
  {2019})},\ \Eprint {http://arxiv.org/abs/1904.09994} {arXiv:1904.09994
  [hep-ph]} \BibitemShut {NoStop}%
\bibitem [{\citenamefont {Song}\ \emph {et~al.}(2021)\citenamefont {Song},
  \citenamefont {Nagorny},\ and\ \citenamefont {Vincent}}]{Song:2021yar}%
  \BibitemOpen
  \bibfield  {author} {\bibinfo {author} {\bibfnamefont {N.}~\bibnamefont
  {Song}}, \bibinfo {author} {\bibfnamefont {S.}~\bibnamefont {Nagorny}}, \
  and\ \bibinfo {author} {\bibfnamefont {A.~C.}\ \bibnamefont {Vincent}},\
  }\href {\doibase 10.1103/PhysRevD.104.103032} {\bibfield  {journal} {\bibinfo
   {journal} {Phys. Rev. D}\ }\textbf {\bibinfo {volume} {104}},\ \bibinfo
  {pages} {103032} (\bibinfo {year} {2021})},\ \Eprint
  {http://arxiv.org/abs/2104.09517} {arXiv:2104.09517 [hep-ph]} \BibitemShut
  {NoStop}%
\bibitem [{\citenamefont {Belanger}\ \emph
  {et~al.}(2022{\natexlab{a}})\citenamefont {Belanger}, \citenamefont
  {Mjallal},\ and\ \citenamefont {Pukhov}}]{Belanger:2021lwd}%
  \BibitemOpen
  \bibfield  {author} {\bibinfo {author} {\bibfnamefont {G.}~\bibnamefont
  {Belanger}}, \bibinfo {author} {\bibfnamefont {A.}~\bibnamefont {Mjallal}}, \
  and\ \bibinfo {author} {\bibfnamefont {A.}~\bibnamefont {Pukhov}},\ }\href
  {\doibase 10.1103/PhysRevD.105.035018} {\bibfield  {journal} {\bibinfo
  {journal} {Phys. Rev. D}\ }\textbf {\bibinfo {volume} {105}},\ \bibinfo
  {pages} {035018} (\bibinfo {year} {2022}{\natexlab{a}})},\ \Eprint
  {http://arxiv.org/abs/2108.08061} {arXiv:2108.08061 [hep-ph]} \BibitemShut
  {NoStop}%
\bibitem [{\citenamefont {Belanger}\ \emph
  {et~al.}(2022{\natexlab{b}})\citenamefont {Belanger}, \citenamefont
  {Mjallal},\ and\ \citenamefont {Pukhov}}]{Belanger:2022qxt}%
  \BibitemOpen
  \bibfield  {author} {\bibinfo {author} {\bibfnamefont {G.}~\bibnamefont
  {Belanger}}, \bibinfo {author} {\bibfnamefont {A.}~\bibnamefont {Mjallal}}, \
  and\ \bibinfo {author} {\bibfnamefont {A.}~\bibnamefont {Pukhov}},\ }\href
  {\doibase 10.1103/PhysRevD.106.095019} {\bibfield  {journal} {\bibinfo
  {journal} {Phys. Rev. D}\ }\textbf {\bibinfo {volume} {106}},\ \bibinfo
  {pages} {095019} (\bibinfo {year} {2022}{\natexlab{b}})},\ \Eprint
  {http://arxiv.org/abs/2205.04101} {arXiv:2205.04101 [hep-ph]} \BibitemShut
  {NoStop}%
\bibitem [{\citenamefont {Cabral-Rosetti}\ \emph {et~al.}(2017)\citenamefont
  {Cabral-Rosetti}, \citenamefont {Gait{\'a}n}, \citenamefont {Montes~de Oca},
  \citenamefont {Osorio~Galicia},\ and\ \citenamefont
  {Garc{\'e}s}}]{Cabral-Rosetti:2017mai}%
  \BibitemOpen
  \bibfield  {author} {\bibinfo {author} {\bibfnamefont {L.~G.}\ \bibnamefont
  {Cabral-Rosetti}}, \bibinfo {author} {\bibfnamefont {R.}~\bibnamefont
  {Gait{\'a}n}}, \bibinfo {author} {\bibfnamefont {J.~H.}\ \bibnamefont
  {Montes~de Oca}}, \bibinfo {author} {\bibfnamefont {R.}~\bibnamefont
  {Osorio~Galicia}}, \ and\ \bibinfo {author} {\bibfnamefont {E.~A.}\
  \bibnamefont {Garc{\'e}s}},\ }\href {\doibase 10.1088/1742-6596/912/1/012047}
  {\bibfield  {journal} {\bibinfo  {journal} {J. Phys. Conf. Ser.}\ }\textbf
  {\bibinfo {volume} {912}},\ \bibinfo {pages} {012047} (\bibinfo {year}
  {2017})}\BibitemShut {NoStop}%
\bibitem [{Sup()}]{SuppMat}%
  \BibitemOpen
  \href@noop {} {}\bibinfo {note} {See Supplemental
  Material~\ref{sec:appendices} for the Boltzmann equations and interaction
  rates governing the asymmetry evolution, additional parameter-space and
  benchmark results, and the effects of the additional quartic
  interactions.}\BibitemShut {Stop}%
\bibitem [{\citenamefont {Bottaro}\ \emph {et~al.}(2022)\citenamefont
  {Bottaro}, \citenamefont {Buttazzo}, \citenamefont {Costa}, \citenamefont
  {Franceschini}, \citenamefont {Panci}, \citenamefont {Redigolo},\ and\
  \citenamefont {Vittorio}}]{Bottaro:2022one}%
  \BibitemOpen
  \bibfield  {author} {\bibinfo {author} {\bibfnamefont {S.}~\bibnamefont
  {Bottaro}}, \bibinfo {author} {\bibfnamefont {D.}~\bibnamefont {Buttazzo}},
  \bibinfo {author} {\bibfnamefont {M.}~\bibnamefont {Costa}}, \bibinfo
  {author} {\bibfnamefont {R.}~\bibnamefont {Franceschini}}, \bibinfo {author}
  {\bibfnamefont {P.}~\bibnamefont {Panci}}, \bibinfo {author} {\bibfnamefont
  {D.}~\bibnamefont {Redigolo}}, \ and\ \bibinfo {author} {\bibfnamefont
  {L.}~\bibnamefont {Vittorio}},\ }\href {\doibase
  10.1140/epjc/s10052-022-10918-5} {\bibfield  {journal} {\bibinfo  {journal}
  {Eur. Phys. J. C}\ }\textbf {\bibinfo {volume} {82}},\ \bibinfo {pages} {992}
  (\bibinfo {year} {2022})},\ \Eprint {http://arxiv.org/abs/2205.04486}
  {arXiv:2205.04486 [hep-ph]} \BibitemShut {NoStop}%
\bibitem [{\citenamefont {Cirelli}\ \emph {et~al.}(2006)\citenamefont
  {Cirelli}, \citenamefont {Fornengo},\ and\ \citenamefont
  {Strumia}}]{Cirelli:2005uq}%
  \BibitemOpen
  \bibfield  {author} {\bibinfo {author} {\bibfnamefont {M.}~\bibnamefont
  {Cirelli}}, \bibinfo {author} {\bibfnamefont {N.}~\bibnamefont {Fornengo}}, \
  and\ \bibinfo {author} {\bibfnamefont {A.}~\bibnamefont {Strumia}},\ }\href
  {\doibase 10.1016/j.nuclphysb.2006.07.012} {\bibfield  {journal} {\bibinfo
  {journal} {Nucl. Phys. B}\ }\textbf {\bibinfo {volume} {753}},\ \bibinfo
  {pages} {178} (\bibinfo {year} {2006})},\ \Eprint
  {http://arxiv.org/abs/hep-ph/0512090} {arXiv:hep-ph/0512090} \BibitemShut
  {NoStop}%
\bibitem [{\citenamefont {Edsjo}\ and\ \citenamefont
  {Gondolo}(1997)}]{Edsjo:1997bg}%
  \BibitemOpen
  \bibfield  {author} {\bibinfo {author} {\bibfnamefont {J.}~\bibnamefont
  {Edsjo}}\ and\ \bibinfo {author} {\bibfnamefont {P.}~\bibnamefont
  {Gondolo}},\ }\href {\doibase 10.1103/PhysRevD.56.1879} {\bibfield  {journal}
  {\bibinfo  {journal} {Phys. Rev. D}\ }\textbf {\bibinfo {volume} {56}},\
  \bibinfo {pages} {1879} (\bibinfo {year} {1997})},\ \Eprint
  {http://arxiv.org/abs/hep-ph/9704361} {arXiv:hep-ph/9704361} \BibitemShut
  {NoStop}%
\bibitem [{\citenamefont {Aghanim}\ \emph {et~al.}(2020)\citenamefont {Aghanim}
  \emph {et~al.}}]{Planck:2018vyg}%
  \BibitemOpen
  \bibfield  {author} {\bibinfo {author} {\bibfnamefont {N.}~\bibnamefont
  {Aghanim}} \emph {et~al.} (\bibinfo {collaboration} {Planck}),\ }\href
  {\doibase 10.1051/0004-6361/201833910} {\bibfield  {journal} {\bibinfo
  {journal} {Astron. Astrophys.}\ }\textbf {\bibinfo {volume} {641}},\ \bibinfo
  {pages} {A6} (\bibinfo {year} {2020})},\ \bibinfo {note} {[Erratum:
  Astron.Astrophys. 652, C4 (2021)]},\ \Eprint
  {http://arxiv.org/abs/1807.06209} {arXiv:1807.06209 [astro-ph.CO]}
  \BibitemShut {NoStop}%
\bibitem [{\citenamefont {Belyaev}\ \emph {et~al.}(2018)\citenamefont
  {Belyaev}, \citenamefont {Cacciapaglia}, \citenamefont {Ivanov},
  \citenamefont {Rojas-Abatte},\ and\ \citenamefont
  {Thomas}}]{Belyaev:2016lok}%
  \BibitemOpen
  \bibfield  {author} {\bibinfo {author} {\bibfnamefont {A.}~\bibnamefont
  {Belyaev}}, \bibinfo {author} {\bibfnamefont {G.}~\bibnamefont
  {Cacciapaglia}}, \bibinfo {author} {\bibfnamefont {I.~P.}\ \bibnamefont
  {Ivanov}}, \bibinfo {author} {\bibfnamefont {F.}~\bibnamefont
  {Rojas-Abatte}}, \ and\ \bibinfo {author} {\bibfnamefont {M.}~\bibnamefont
  {Thomas}},\ }\href {\doibase 10.1103/PhysRevD.97.035011} {\bibfield
  {journal} {\bibinfo  {journal} {Phys. Rev. D}\ }\textbf {\bibinfo {volume}
  {97}},\ \bibinfo {pages} {035011} (\bibinfo {year} {2018})},\ \Eprint
  {http://arxiv.org/abs/1612.00511} {arXiv:1612.00511 [hep-ph]} \BibitemShut
  {NoStop}%
\bibitem [{\citenamefont {Blinov}\ \emph {et~al.}(2016)\citenamefont {Blinov},
  \citenamefont {Kozaczuk}, \citenamefont {Morrissey},\ and\ \citenamefont
  {de~la Puente}}]{Blinov:2015qva}%
  \BibitemOpen
  \bibfield  {author} {\bibinfo {author} {\bibfnamefont {N.}~\bibnamefont
  {Blinov}}, \bibinfo {author} {\bibfnamefont {J.}~\bibnamefont {Kozaczuk}},
  \bibinfo {author} {\bibfnamefont {D.~E.}\ \bibnamefont {Morrissey}}, \ and\
  \bibinfo {author} {\bibfnamefont {A.}~\bibnamefont {de~la Puente}},\ }\href
  {\doibase 10.1103/PhysRevD.93.035020} {\bibfield  {journal} {\bibinfo
  {journal} {Phys. Rev. D}\ }\textbf {\bibinfo {volume} {93}},\ \bibinfo
  {pages} {035020} (\bibinfo {year} {2016})},\ \Eprint
  {http://arxiv.org/abs/1510.08069} {arXiv:1510.08069 [hep-ph]} \BibitemShut
  {NoStop}%
\bibitem [{\citenamefont {Aad}\ \emph {et~al.}(2022)\citenamefont {Aad} \emph
  {et~al.}}]{ATLAS:2022rme}%
  \BibitemOpen
  \bibfield  {author} {\bibinfo {author} {\bibfnamefont {G.}~\bibnamefont
  {Aad}} \emph {et~al.} (\bibinfo {collaboration} {ATLAS}),\ }\href {\doibase
  10.1140/epjc/s10052-022-10489-5} {\bibfield  {journal} {\bibinfo  {journal}
  {Eur. Phys. J. C}\ }\textbf {\bibinfo {volume} {82}},\ \bibinfo {pages} {606}
  (\bibinfo {year} {2022})},\ \Eprint {http://arxiv.org/abs/2201.02472}
  {arXiv:2201.02472 [hep-ex]} \BibitemShut {NoStop}%
\bibitem [{\citenamefont {Aad}\ \emph {et~al.}(2024)\citenamefont {Aad} \emph
  {et~al.}}]{ATLAS:2024umc}%
  \BibitemOpen
  \bibfield  {author} {\bibinfo {author} {\bibfnamefont {G.}~\bibnamefont
  {Aad}} \emph {et~al.} (\bibinfo {collaboration} {ATLAS}),\ }\href {\doibase
  10.1103/PhysRevLett.132.221801} {\bibfield  {journal} {\bibinfo  {journal}
  {Phys. Rev. Lett.}\ }\textbf {\bibinfo {volume} {132}},\ \bibinfo {pages}
  {221801} (\bibinfo {year} {2024})},\ \Eprint
  {http://arxiv.org/abs/2401.14046} {arXiv:2401.14046 [hep-ex]} \BibitemShut
  {NoStop}%
\bibitem [{\citenamefont {Aad}\ \emph {et~al.}(2025)\citenamefont {Aad} \emph
  {et~al.}}]{ATLAS:2025lhc}%
  \BibitemOpen
  \bibfield  {author} {\bibinfo {author} {\bibfnamefont {G.}~\bibnamefont
  {Aad}} \emph {et~al.} (\bibinfo {collaboration} {ATLAS}),\ }\href@noop {} {\
  (\bibinfo {year} {2025})},\ \Eprint {http://arxiv.org/abs/2511.20042}
  {arXiv:2511.20042 [hep-ex]} \BibitemShut {NoStop}%
\bibitem [{\citenamefont {Hayrapetyan}\ \emph
  {et~al.}(2025{\natexlab{a}})\citenamefont {Hayrapetyan} \emph
  {et~al.}}]{CMS:2025qkk}%
  \BibitemOpen
  \bibfield  {author} {\bibinfo {author} {\bibfnamefont {A.}~\bibnamefont
  {Hayrapetyan}} \emph {et~al.} (\bibinfo {collaboration} {CMS}),\ }\href@noop
  {} {\  (\bibinfo {year} {2025}{\natexlab{a}})},\ \Eprint
  {http://arxiv.org/abs/2511.08212} {arXiv:2511.08212 [hep-ex]} \BibitemShut
  {NoStop}%
\bibitem [{\citenamefont {Hayrapetyan}\ \emph
  {et~al.}(2025{\natexlab{b}})\citenamefont {Hayrapetyan} \emph
  {et~al.}}]{CMS:2025mie}%
  \BibitemOpen
  \bibfield  {author} {\bibinfo {author} {\bibfnamefont {A.}~\bibnamefont
  {Hayrapetyan}} \emph {et~al.} (\bibinfo {collaboration} {CMS}),\ }\href@noop
  {} {\  (\bibinfo {year} {2025}{\natexlab{b}})},\ \Eprint
  {http://arxiv.org/abs/2511.16394} {arXiv:2511.16394 [hep-ex]} \BibitemShut
  {NoStop}%
\bibitem [{\citenamefont {Belyaev}\ \emph {et~al.}(2021)\citenamefont
  {Belyaev}, \citenamefont {Prestel}, \citenamefont {Rojas-Abbate},\ and\
  \citenamefont {Zurita}}]{Belyaev:2020wok}%
  \BibitemOpen
  \bibfield  {author} {\bibinfo {author} {\bibfnamefont {A.}~\bibnamefont
  {Belyaev}}, \bibinfo {author} {\bibfnamefont {S.}~\bibnamefont {Prestel}},
  \bibinfo {author} {\bibfnamefont {F.}~\bibnamefont {Rojas-Abbate}}, \ and\
  \bibinfo {author} {\bibfnamefont {J.}~\bibnamefont {Zurita}},\ }\href
  {\doibase 10.1103/PhysRevD.103.095006} {\bibfield  {journal} {\bibinfo
  {journal} {Phys. Rev. D}\ }\textbf {\bibinfo {volume} {103}},\ \bibinfo
  {pages} {095006} (\bibinfo {year} {2021})},\ \Eprint
  {http://arxiv.org/abs/2008.08581} {arXiv:2008.08581 [hep-ph]} \BibitemShut
  {NoStop}%
\bibitem [{\citenamefont {Aalbers}\ \emph {et~al.}(2024)\citenamefont {Aalbers}
  \emph {et~al.}}]{XLZD:2024nsu}%
  \BibitemOpen
  \bibfield  {author} {\bibinfo {author} {\bibfnamefont {J.}~\bibnamefont
  {Aalbers}} \emph {et~al.} (\bibinfo {collaboration} {XLZD}),\ }\href@noop {}
  {\  (\bibinfo {year} {2024})},\ \Eprint {http://arxiv.org/abs/2410.17137}
  {arXiv:2410.17137 [hep-ex]} \BibitemShut {NoStop}%
\bibitem [{\citenamefont {Abdukerim}\ \emph {et~al.}(2025)\citenamefont
  {Abdukerim} \emph {et~al.}}]{PANDA-X:2024dlo}%
  \BibitemOpen
  \bibfield  {author} {\bibinfo {author} {\bibfnamefont {A.}~\bibnamefont
  {Abdukerim}} \emph {et~al.} (\bibinfo {collaboration} {PANDA-X, PandaX}),\
  }\href {\doibase 10.1007/s11433-024-2539-y} {\bibfield  {journal} {\bibinfo
  {journal} {Sci. China Phys. Mech. Astron.}\ }\textbf {\bibinfo {volume}
  {68}},\ \bibinfo {pages} {221011} (\bibinfo {year} {2025})},\ \Eprint
  {http://arxiv.org/abs/2402.03596} {arXiv:2402.03596 [hep-ex]} \BibitemShut
  {NoStop}%
\bibitem [{\citenamefont {Hayrapetyan}\ \emph {et~al.}(2024)\citenamefont
  {Hayrapetyan} \emph {et~al.}}]{CMS:2023mny}%
  \BibitemOpen
  \bibfield  {author} {\bibinfo {author} {\bibfnamefont {A.}~\bibnamefont
  {Hayrapetyan}} \emph {et~al.} (\bibinfo {collaboration} {CMS}),\ }\href
  {\doibase 10.1103/PhysRevD.109.072007} {\bibfield  {journal} {\bibinfo
  {journal} {Phys. Rev. D}\ }\textbf {\bibinfo {volume} {109}},\ \bibinfo
  {pages} {072007} (\bibinfo {year} {2024})},\ \Eprint
  {http://arxiv.org/abs/2309.16823} {arXiv:2309.16823 [hep-ex]} \BibitemShut
  {NoStop}%
\bibitem [{\citenamefont {de~Blas}\ \emph {et~al.}(2025)\citenamefont {de~Blas}
  \emph {et~al.}}]{deBlas:2025gyz}%
  \BibitemOpen
  \bibfield  {author} {\bibinfo {author} {\bibfnamefont {J.}~\bibnamefont
  {de~Blas}} \emph {et~al.},\ }\href {\doibase 10.17181/CERN.35CH.2O2P} {\
  (\bibinfo {year} {2025}),\ 10.17181/CERN.35CH.2O2P},\ \Eprint
  {http://arxiv.org/abs/2511.03883} {arXiv:2511.03883 [hep-ex]} \BibitemShut
  {NoStop}%
\end{thebibliography}%

\newpage
\onecolumngrid

\appendix

\section{Supplemental Material to Keeping It Renormalizable:\\
Minimal Baryogenesis induced Asymmetric Dark Matter}\label{sec:appendices}
\begin{center}
\textit{Authors:} Miguel Escudero Abenza, Thomas Hambye and Chandan Hati
\end{center}

\begin{center}
\textbf{Boltzmann Equations \& Parameter space}
\end{center}

In this Supplemental Material we display the various Boltzmann equations that are used to solve for the particle asymmetry evolution. To this end, we also calculate the rates for all relevant processes. After these equations are presented, we show figures displaying the various regions of the parameter space and highlights the dynamics. Finally, we have a small section that shows that the additional quartic couplings which can modify the asymmetry evolution and which are ignored in the main text have a very minor impact on our results.

\subsection*{Interactions and their rates}

To determine the evolution of the $H$, $H'$ and $\phi$ number density asymmetries, one must first compute the cross section of the various processes that can change them. These processes are\footnote{In principle, the processes $H' \leftrightarrow \bar{\phi} \bar{\phi} H$ and $H \leftrightarrow \phi \phi H'$ are possible kinematically (particularly in the presence of strong thermal effects) but since in vacuum we work in a regime where $m_\phi >m_{H'}$ their impact is most likely negligible, and we ignore them in our calculation.}:
\begin{align}
     H' H'  &\leftrightarrow H H  \,,\quad  \bar{H'} \bar{H'}  \leftrightarrow \bar{H} \bar{H}   \,, \quad H'  \bar{H}  \leftrightarrow \bar{H'} H  \,,\quad   \bar{H'}  H  \leftrightarrow H' \bar{H}  \quad \text{(induced by $\lambda_5$)}\\
         \phi \phi &\leftrightarrow H \bar{H}' \,,\quad \,\,\,\,\, \bar{\phi} \bar{\phi} \leftrightarrow \bar{H} {H}'\,,\quad \,\,   \phi  \bar{H} \leftrightarrow \bar{\phi} \bar{H}'\,,\quad  \,\,\,\,\,\,\bar{\phi}  H \leftrightarrow \phi {H}' \quad\,\, \text{(induced by $\lambda_{\phi^2 HH'}$)}
\end{align}

Before electroweak symmetry breaking our interactions always involve four-point scalar vertices and we can easily calculate all cross sections and their thermal averages. The general formulas are:
\begin{align}
    \sigma_{1+2\to 3+4}  &= \frac{S|\mathcal{M}|^2}{16\pi s} \frac{\lambda^{1/2}(s,m_3^2,m_4^2)}{\lambda^{1/2}(s,m_1^2,m_2^2)}\,,
\end{align}
where here we have assumed that $|\mathcal{M}|^2$ is constant (as for our cases of interest), and where $\lambda(x,y,z) = x^2+y^2+z^2-2xy-2xz-2yz $ is the Kallen function. Furthermore, $S$ is the symmetry factor that is $1/2$ if there are two identical particles in the final state, and unity otherwise. 

The thermally averaged cross section for a process $1+2 \to 3+ 4$ reads~\cite{Edsjo:1997bg}:
\begin{align}
\left< \sigma v\right> &= \frac{1}{8 T m_1^2 m_2^2 K_2(m_1/T) K_2(m_2/T)} \int_{{\rm max}[(m_1+m_2)^2,(m_3+m_4)^2]}^{\infty} ds \, s^{3/2} K_1\left(\frac{\sqrt{s}}{T} \right) \lambda \left[1,\frac{m_1^2}{s},\frac{m_2^2}{s}\right] \sigma(s)\,,
 \end{align}
where $K_1$ and $K_2$ are modified Bessel functions of the second kind of order 1 and 2, respectively. 

The two limits of interest are the case where $T\gg m_{i}$ and the case where $m_1\simeq m_2 \gg m_{3,4}$ and $T\ll m_1$. In those cases one gets:
\begin{align}
    \left<\sigma v\right>|_{m_i = 0} &= S\frac{|\mathcal{M}|^2 }{32 \pi} \frac{1}{4 T^2}\,, \quad ({T\gg m})\\
    \left<\sigma v\right>|_{m_{1}=m_2 = m ,\,m_{3,4}= 0\,} &= S\frac{|\mathcal{M}|^2 }{32\pi} \frac{1}{m^2}\,.\quad \,\,\,({T\ll m})
\end{align}
It is convenient to interpolate these rates between these two regimes, and we find: 
\begin{subequations}\label{eq:Rates}
\begin{align}
    \left<\sigma v({H' H'  \leftrightarrow H H })\right> &\simeq \frac{3}{2}  \frac{|\lambda_5|^2}{32\pi} \left[\frac{1}{m_{H'}^2+ 2.5m_{H'} T +4 T^2}\right] \,,\\
    \left<\sigma v(H'  \bar{H}  \leftrightarrow \bar{H'} H)\right> &\simeq  3 \frac{|\lambda_5|^2}{32\pi} \left[\frac{1}{m_{H'}^2+ 2.5m_{H'} T +4 T^2}\right] \,,\\
    \left<\sigma v (\phi \phi \leftrightarrow {H}   \bar{H'})\right> &\simeq 2\frac{|\lambda_{\phi^2HH'}|^2}{32\pi} \left[\frac{1}{m_{\phi}^2+ 2.5m_{\phi} T +4 T^2}\right]\,,\\
    \left<\sigma v (\phi {\bar H} \leftrightarrow  \bar{\phi}  \bar{H'})\right> &\simeq \frac{|\lambda_{\phi^2HH'}|^2}{32\pi} \left[\frac{1}{m_{\phi}^2+ 2.5m_{\phi} T +4 T^2}\right]\,.
\end{align}
\end{subequations}
We obtained these rates by comparing them against the explicit numerical integration and we find they are accurate to better than 10\% across the relevant temperature regime for our study. The pre-factors in from of each formula corresponds to the average over initial states and sum over final states, as required for the Boltzmann equations below.

\subsection*{The Boltzmann Equations}

Having identified the processes that provide a transfer of particle asymmetries in the early Universe we can proceed with writing down the Boltzmann equations for our scenario. For simplicity, we work assuming that all particles follow Maxwell-Boltzmann statistics because it simplifies greatly the collision terms and makes nice detailed balance relations, and also because it should be accurate to $\sim 10\%$. In this context, we define the yields as $Y_i \equiv n_i/s$, and $Y_{i}^{\rm eq} = n_i^{\rm eq}/s$. Furthermore, considering $\Delta H' \equiv Y_{H'} - Y_{\bar H'}$, $\Delta H \equiv Y_{H} - Y_{\bar H}$ and $\Delta \phi \equiv Y_{\phi} - Y_{\bar \phi}$, we get:
\begin{subequations}
\begin{align}
\!\!\!\!\!    \frac{d \Delta_{H'}}{d x} = -\frac{s}{H x} \bigg\{ & + \left<\sigma v (H'H'\to H H) \right> Y_{H'}^{\rm eq}  Y_{H'}^{\rm eq} \left[\frac{Y_{H'}}{Y_{H'}^{\rm eq}} \frac{Y_{H'}}{Y_{H'}^{\rm eq}} - \frac{Y_H}{Y_H^{\rm eq}} \frac{Y_H}{Y_H^{\rm eq}} - \left(\frac{Y_{\bar{H}'}}{Y_{{H}'}^{\rm eq}} \frac{Y_{\bar{H}'}}{Y_{H'}^{\rm eq}} - \frac{Y_{\bar H}}{Y_H^{\rm eq}} \frac{Y_{\bar H}}{Y_H^{\rm eq}}\right) \right]  \quad {(\lambda_5)}\\ 
     &+2\left<\sigma v (H'\bar{H}\to \bar{H'} H) \right> Y_{H'}^{\rm eq}  Y_{H}^{\rm eq} \left[\frac{Y_{H'}}{Y_{H'}^{\rm eq}} \frac{Y_{\bar{H}}}{Y_{H}^{\rm eq}} - \frac{Y_{\bar{H}'}}{Y_{H'}^{\rm eq}} \frac{Y_H}{Y_H^{\rm eq}} \right]  \quad {(\lambda_5)} \\
    &-\frac{1}{2}\left<\sigma v (\bar{\phi}\bar{\phi} \to \bar{H} H'  ) \right> Y_{\phi}^{\rm eq}  Y_{\phi}^{\rm eq} \left[\frac{Y_{\bar{\phi}}}{Y_{\phi}^{\rm eq}} \frac{Y_{\bar{\phi}}}{Y_{\phi}^{\rm eq}}-\frac{Y_{\bar{H}}}{Y_{H}^{\rm eq}} \frac{Y_{H'}}{Y_{H'}^{\rm eq}} -\left( \frac{Y_{{\phi}}}{Y_{\phi}^{\rm eq}} \frac{Y_{{\phi}}}{Y_{\phi}^{\rm eq}}-\frac{Y_{{H}}}{Y_{H}^{\rm eq}} \frac{Y_{\bar{H}'}}{Y_{H'}^{\rm eq}}\right)\right] \quad (\lambda_{\phi^2HH'}) \\
    &-\left<\sigma v (\bar{\phi} H \to {\phi} H') \right> Y_{\phi}^{\rm eq}  Y_{H}^{\rm eq} \left[\frac{Y_{\bar{\phi}}}{Y_{\phi}^{\rm eq}} \frac{Y_{H}}{Y_{H}^{\rm eq}} - \frac{Y_{{\phi}}}{Y_{\phi}^{\rm eq}} \frac{Y_{H'}}{Y_{H'}^{\rm eq}} - \left(\frac{Y_{{\phi}}}{Y_{\phi}^{\rm eq}} \frac{Y_{\bar H}}{Y_{H}^{\rm eq}} - \frac{Y_{\bar{\phi}}}{Y_{\phi}^{\rm eq}} \frac{Y_{\bar H'}}{Y_{H'}^{\rm eq}}  \right)\right]\bigg\} \,,  \quad (\lambda_{\phi^2HH'}) 
\end{align}
\end{subequations}
\begin{subequations}\label{eq:dDeltaphidx_complete}
\begin{align}
    \frac{d\Delta_\phi}{dx} =   -\frac{s}{H x} \bigg\{ 
    &-2\left<\sigma v (\bar{\phi} H \to {\phi} H') \right> Y_{\phi}^{\rm eq}  Y_{H}^{\rm eq} \left[\frac{Y_{\bar{\phi}}}{Y_{\phi}^{\rm eq}} \frac{Y_{H}}{Y_{H}^{\rm eq}} - \frac{Y_{{\phi}}}{Y_{\phi}^{\rm eq}} \frac{Y_{H'}}{Y_{H'}^{\rm eq}} - \left(\frac{Y_{{\phi}}}{Y_{\phi}^{\rm eq}} \frac{Y_{\bar H}}{Y_{H}^{\rm eq}} - \frac{Y_{{\bar \phi}}}{Y_{\phi}^{\rm eq}} \frac{Y_{\bar H'}}{Y_{H'}^{\rm eq}} \right)  \right]  \quad (\lambda_{\phi^2HH'}) \\
    &+\left<\sigma v ({\phi}{\phi} \to H \bar{H}' ) \right> Y_{\phi}^{\rm eq}  Y_{\phi}^{\rm eq} \left[\frac{Y_{{\phi}}}{Y_{\phi}^{\rm eq}} \frac{Y_{{\phi}}}{Y_{\phi}^{\rm eq}}-\frac{Y_{{H}}}{Y_{H}^{\rm eq}} \frac{Y_{\bar H'}}{Y_{H'}^{\rm eq}} -\left(\frac{Y_{{\bar \phi}}}{Y_{\phi}^{\rm eq}} \frac{Y_{{\bar \phi}}}{Y_{\phi}^{\rm eq}}-\frac{Y_{{\bar H}}}{Y_{H}^{\rm eq}} \frac{Y_{ H'}}{Y_{H'}^{\rm eq}}\right) \right] \bigg\} \,. \quad (\lambda_{{\phi}^2HH'}) 
\end{align}
\end{subequations}
where the ``$H$'' appearing in the denominator multiplying the brackets stands for the Hubble rate, and 
\begin{align}\label{eq:HSMevol}
    \frac{d\Delta_H}{dx} = - \frac{13}{79}\frac{d \Delta_{H'}}{dx}\,. \quad \text{($B-L$ and $Q$ conservation and all SM interactions active)}
\end{align}
In all these equations we have indicated the origin of each of the terms and the relevant rates are highlighted above. 

The equation for the SM Higgs doublet arises from the conservation of $B-L$ and electric charge $Q$. These chemical potentials in our scenario read~\cite{Harvey:1990qw}:
\begin{align}
    Q &= 2\mu_H + 4\mu_H N_f + 8 N_f \mu_{u_L}  {+2 \mu_{H'}}\,,\\
    B-L &= N_f (\mu_{H}+13\mu_{u_L})\,,
\end{align}
where here $\mu_{u_L}$ is the chemical potential of left-handed $u$-quarks, $\mu_{H}$ is the chemical potential of the SM Higgs doublet, and $\mu_{H'}$ the one of the inert doublet. In these expressions we have imposed the $SU(2)_L$ $T_3$ charge to vanish, $Q_3 = 0$, which implies $\mu_W = 0$. $N_f = 3$ is the number of generations in the SM. In our scenario there is no BSM source of $B-L$ charge and we also know that $Q = 0$. As such, imposing $dQ/dt = 0$ and $d(B-L)/dt = 0$ relates the derivatives of the two Higgs chemical potentials and tells us that:
\begin{align}
   \frac{d \mu_{H}}{dt} = - \frac{13}{79} \frac{d \mu_{H'} }{dt} \,,
\end{align}
which is the equation we showed above (since both $H$ and $H'$ are effectively relativistic at the temperatures of interest). 

Our starting conditions for the system at very high temperatures are: 
\begin{align}
    \Delta H' = 0\,,\quad \Delta \phi = 0\,,\quad \mu_{H} = - B/7\,,
\end{align}
where for the last term we take into account the relationship between the chemical potentials of baryons in the Standard Model and the Higgs doublet at $T>T_{\rm EW}\simeq 160\,{\rm GeV}$~\cite{Harvey:1990qw}. For the baryon asymmetry yield we take the observed value ($Y_B = 8.674\times 10^{-11}$ as implied by Planck CMB observations~\cite{Planck:2018vyg}), which is related to the baryon chemical potential as  $B = 4\pi^2 g_{\star S} Y_B/15$. The number of relativistic degrees of freedom which holds is $g_{\star S} = 106$, since for the  temperature range of interest all SM species are in thermal equilibrium.

Technically, the equations for $\Delta H'$ and $\Delta \phi$ depend upon their individual yields, but since most of our dynamics happen at $T> m/20$ we can simply consider $Y_i = Y_i^{\rm eq} + \Delta_i/2$ and $Y_{\bar i} = Y_i^{\rm eq} - \Delta_i/2$. This allows us to readily linearize all our equations which under this approximation then become:
\begin{subequations}
\begin{align}
\!\!\!\!\!    \frac{d \Delta_{H'}}{d x} &= -\frac{s}{H x} \bigg\{  +2\big(\left<\sigma v (H'H'\to H H) \right>Y_{H'}^{\rm eq}  + \left<\sigma v (H'\bar{H}\to \bar{H'} H) \right> Y_{H}^{\rm eq} \big) \left[ \Delta H' - \frac{Y_{H'}^{\rm eq}}{Y_{H}^{\rm eq}} \Delta H \right]  \\ 
    &+\big(\frac{1}{2}\left<\sigma v ({\phi}{\phi} \to {H} \bar{H}'  ) \right>   Y_{\phi}^{\rm eq} + \left<\sigma v ({\phi} \bar{H} \to \bar{\phi} \bar{H}') \right>Y_{H}^{\rm eq} \big) \left[2\Delta \phi + \frac{Y_\phi^{\rm eq}}{Y_{H'}^{\rm eq}}\Delta H' - \frac{Y_\phi^{\rm eq}}{Y_{H}^{\rm eq}}\Delta H\right]\bigg\} 
\end{align}
\end{subequations}
\begin{align}\label{eq:DeltaPhieasy}
    \frac{d\Delta_\phi}{dx} =   -\frac{s}{H x} \bigg\{ +\big(2\left<\sigma v ({\phi} \bar{H} \to \bar{\phi} \bar{H}') \right> Y_{H}^{\rm eq}+\left<\sigma v ({\phi}{\phi} \to H \bar{H}' ) \right> Y_{\phi}^{\rm eq}\big) \left[2\Delta \phi + \frac{Y_\phi^{\rm eq}}{Y_{H'}^{\rm eq}}\Delta H' - \frac{Y_\phi^{\rm eq}}{Y_{H}^{\rm eq}}\Delta H\right] \bigg\} \,.
\end{align}
These equations, together
 with Eq.~\eqref{eq:HSMevol}, form a closed system of equations. Finally, we note that 
\begin{align}
    Y_i^{\rm eq}  \equiv n_i^{\rm eq}/s = \frac{g_i}{\pi^2} \frac{1}{2\pi^2 g_{\star S}/45} \frac{(\frac{m_i}{T})^2 K_2(\frac{m_i}{T})}{2}\,,
\end{align}
where $g_\phi = 1$, $g_{H} = g_{H'} = 2$. To solve the equations above we use $x = m_{H'}/T$. For all of our results, we solve the evolution of asymmetries down to $T = T_{\rm EW}$. At $T< T_{\rm EW}$, $\Delta H' = \Delta H = 0$ and therefore the $\Delta \phi $ asymmetry may only be washed out. We explicitly checked that solving for $T\ll T_{\rm EW}$ and taking this washout into account only changes slightly the upper branch of the solutions in the $\lambda_{\phi^2 HH'}$-$m_\phi$ plane, as in Fig.~\ref{fig:Mechanism_results}. 

We note that Eq.~\eqref{eq:simple_phi_equation} in the main text has been obtained from Eq.~\eqref{eq:DeltaPhieasy} by taking the ultrarelativistic limit for $Y_i$ and for the scattering cross sections, see Eq.~\eqref{eq:Rates}.

To conclude, we highlight that the presence of a thermalized $H'$ or $\phi$ does not influence significantly the primordial $B-L$ asymmetry needed to explain the baryon asymmetry of the Universe. At $T>T_{\rm EW}$ we have:
\begin{align}
    B &= \frac{28}{79} (B-L) \simeq 0.354\,(B-L)  \,,\quad \text{SM}\\
    B &= \frac{8}{23} (B-L)\simeq 0.348\,(B-L)\,,\quad \text{SM}+H'\text{ in equilibrium}
\end{align}
so that the difference is 2\%, which is effectively irrelevant for any baryogenesis mechanism.

\begin{figure*}[hbtp]
\centering
\begin{tabular}{cc}
\includegraphics[width=0.33\textwidth]{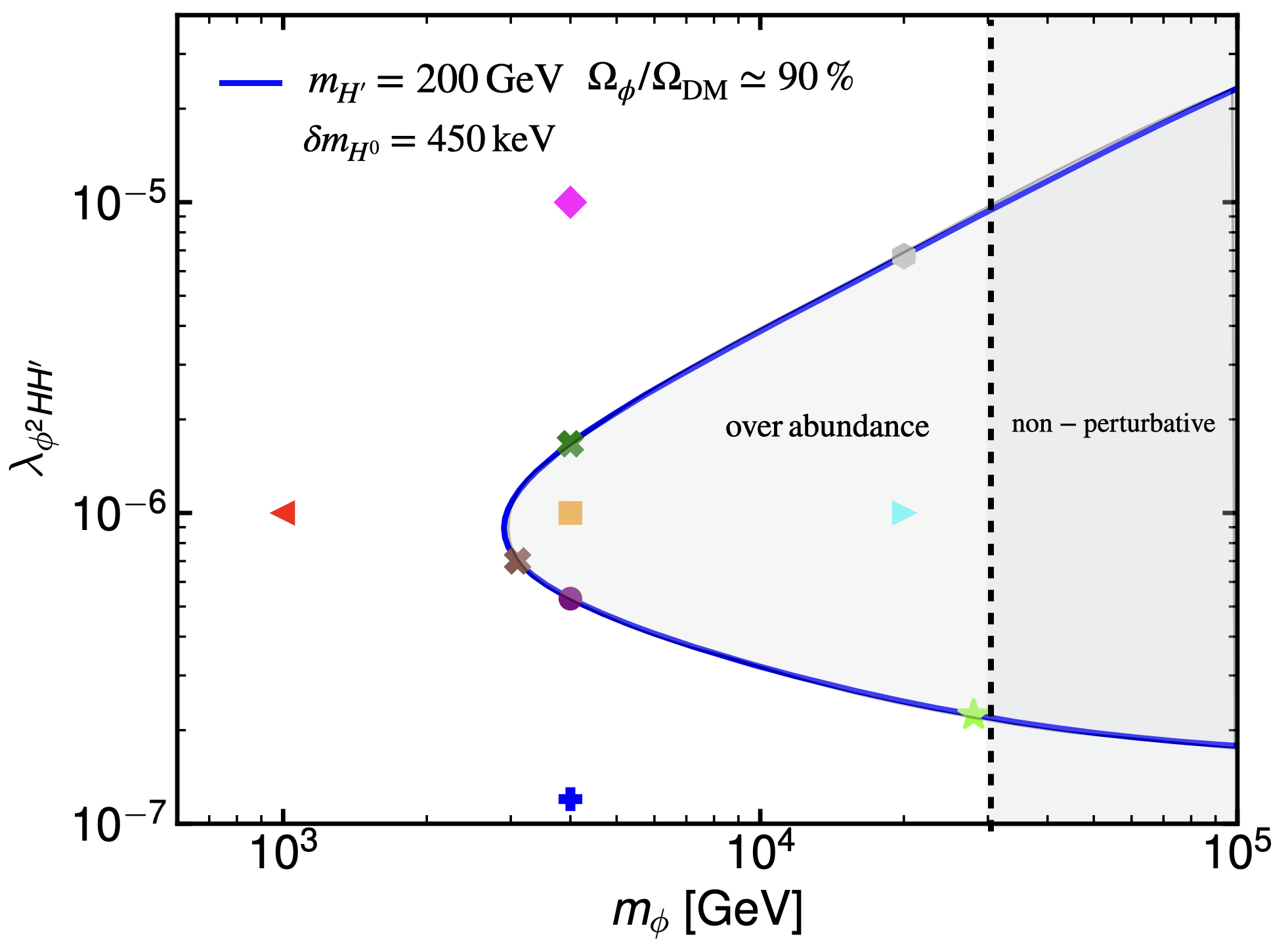} & \includegraphics[width=0.33\textwidth]{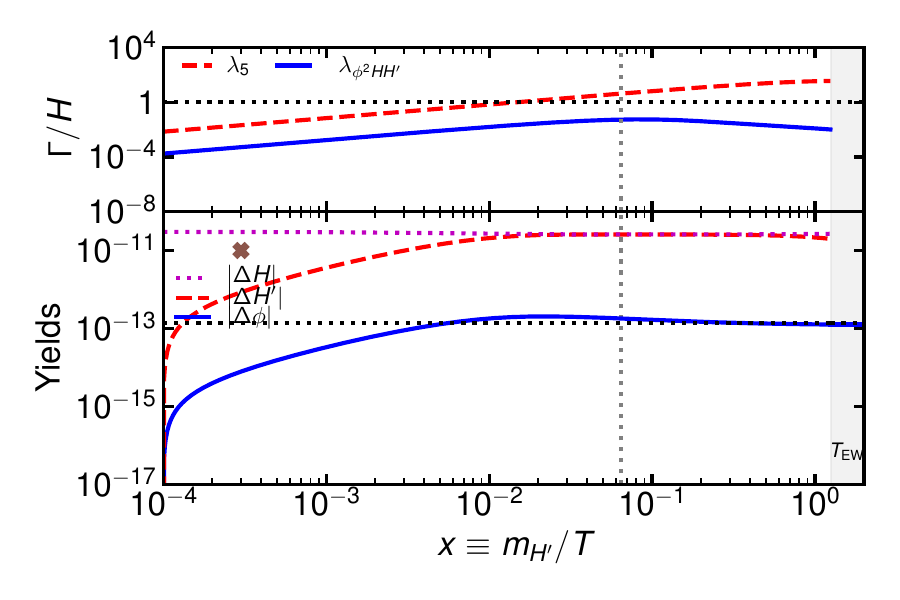} 
\end{tabular}
\begin{tabular}{ccc}
\includegraphics[width=0.33\textwidth]{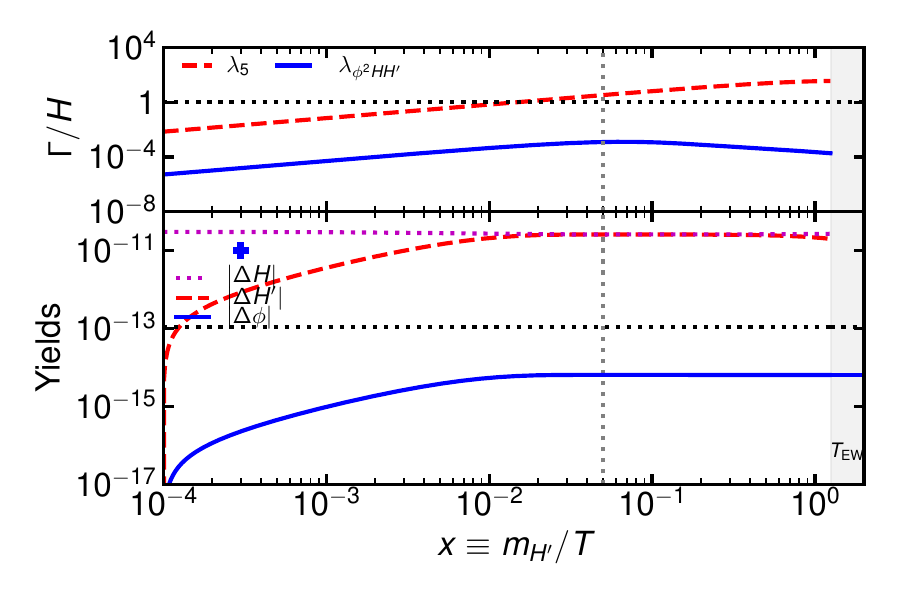} & \includegraphics[width=0.33\textwidth]{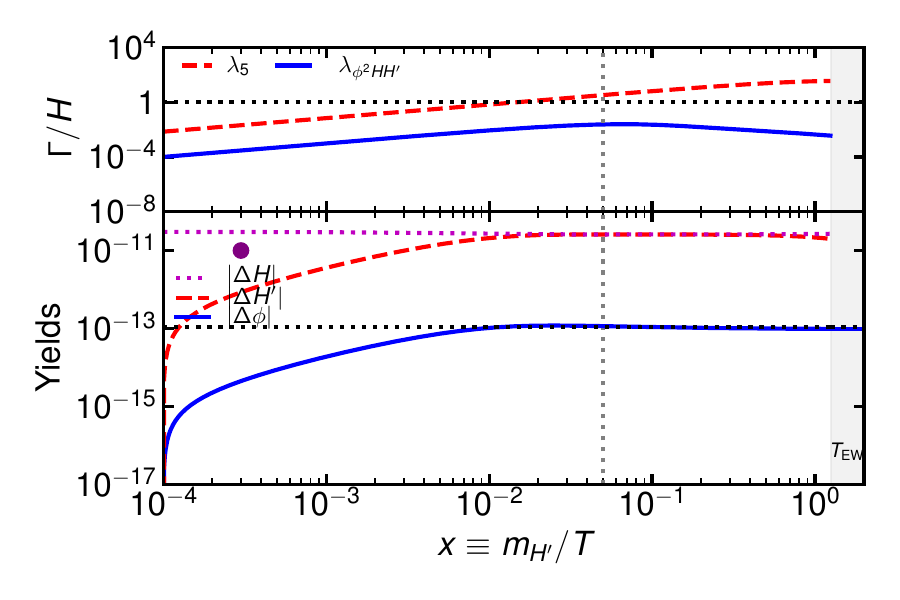} & \includegraphics[width=0.33\textwidth]{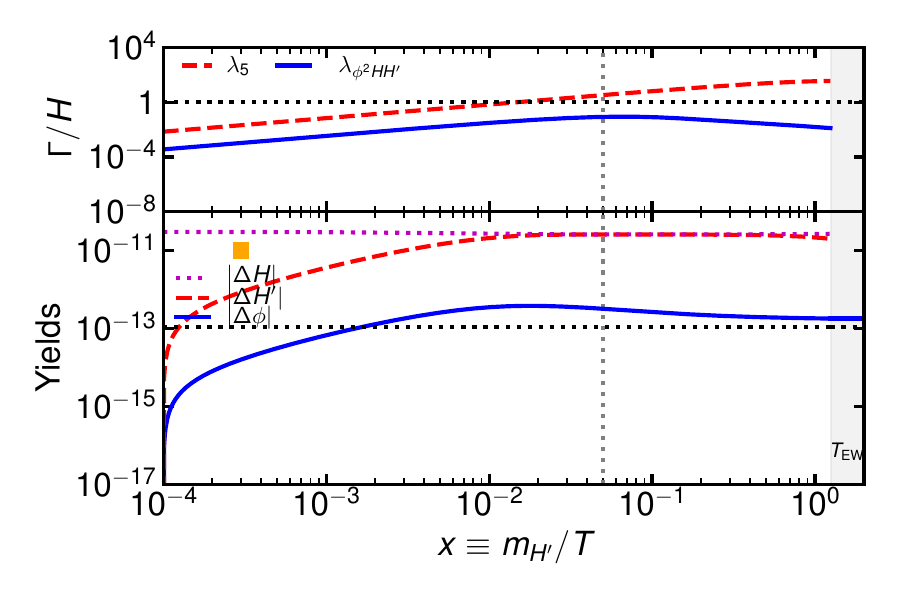} \\
\includegraphics[width=0.33\textwidth]{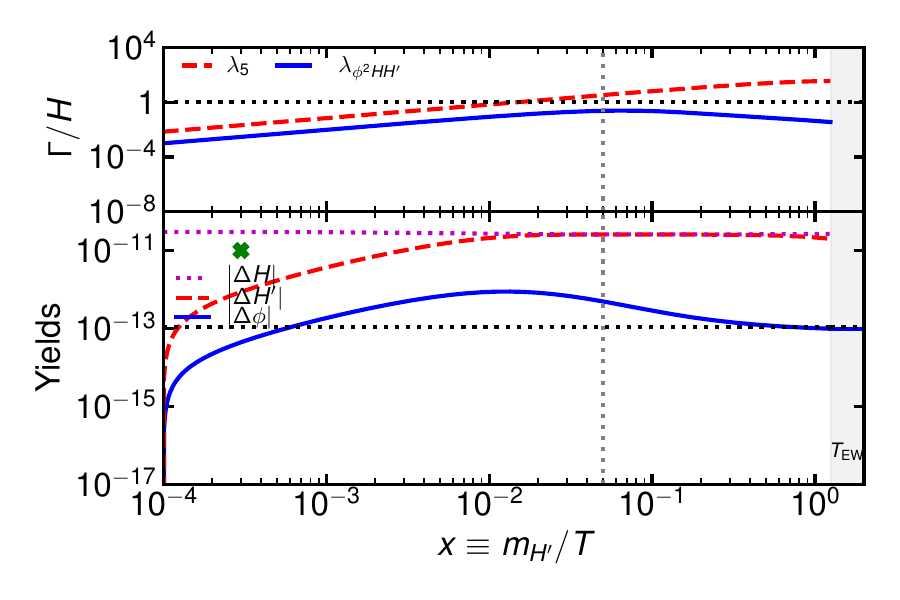} & \includegraphics[width=0.33\textwidth]{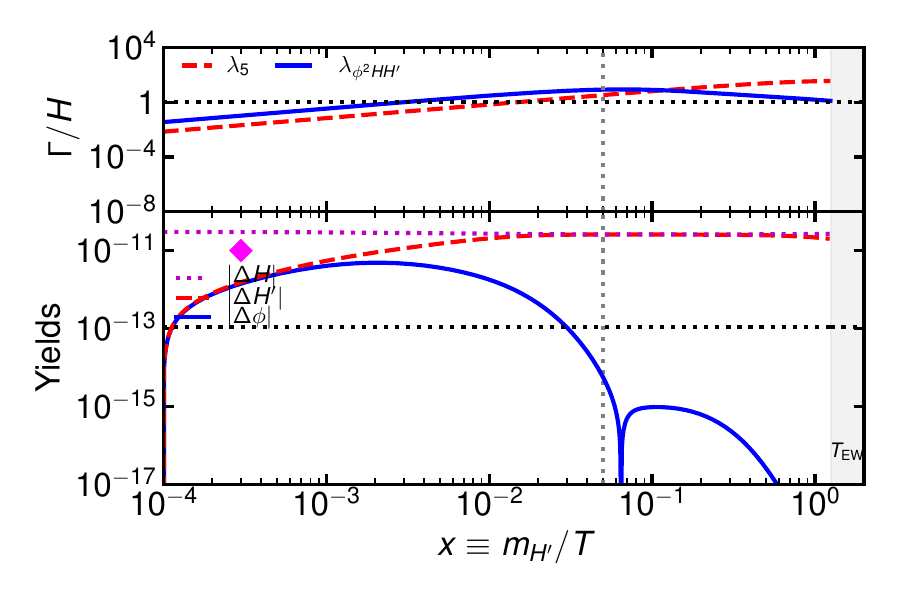} &  \includegraphics[width=0.33\textwidth]{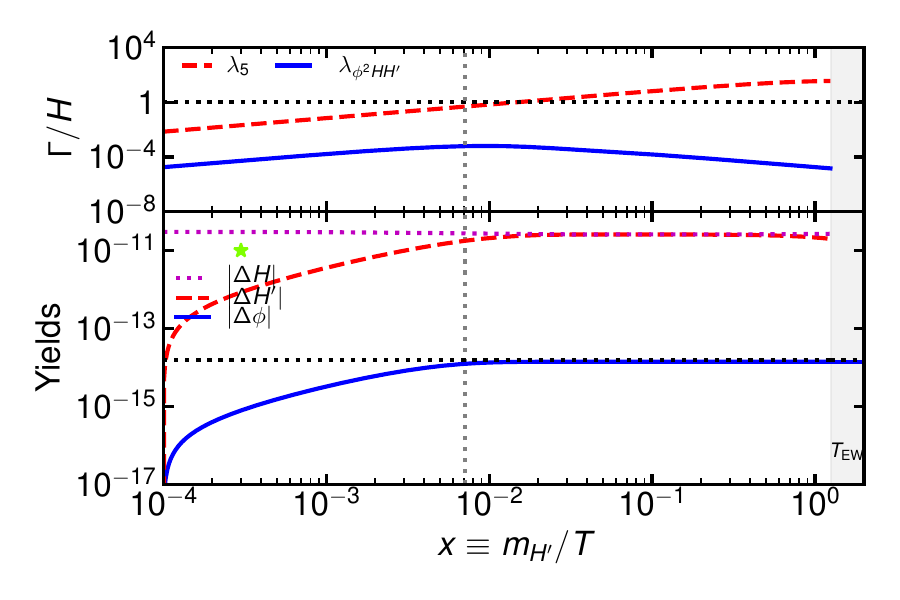} \\
 \includegraphics[width=0.33\textwidth]{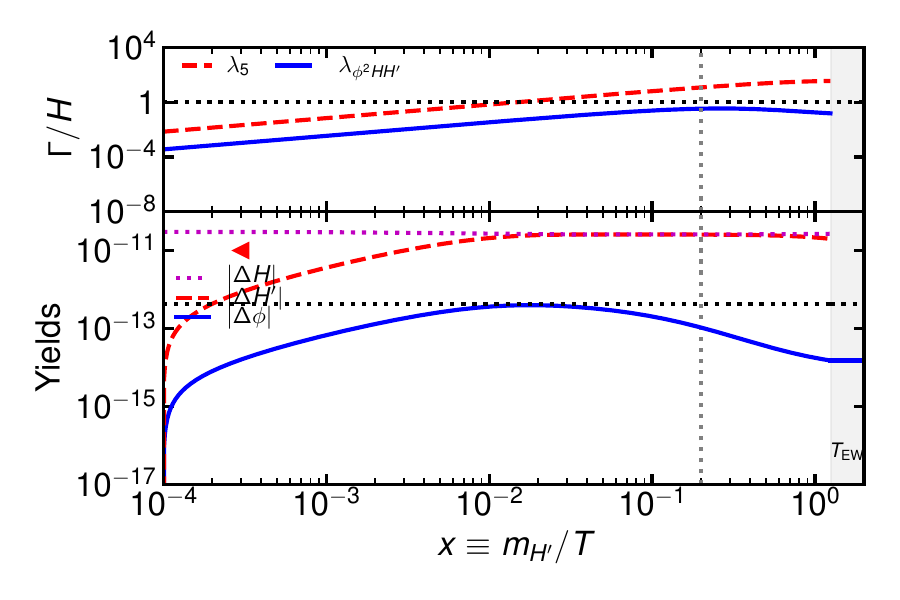} & \includegraphics[width=0.33\textwidth]{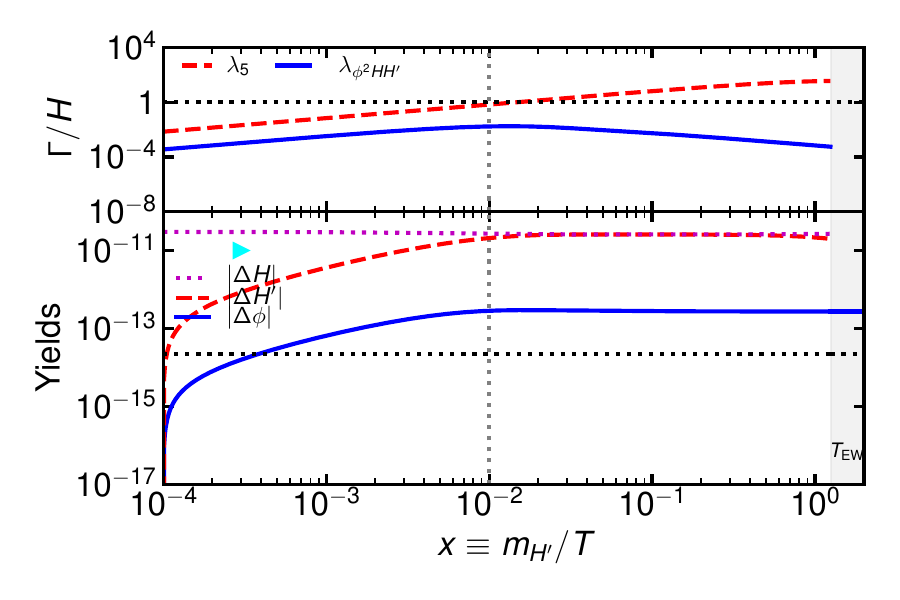} & \includegraphics[width=0.33\textwidth]{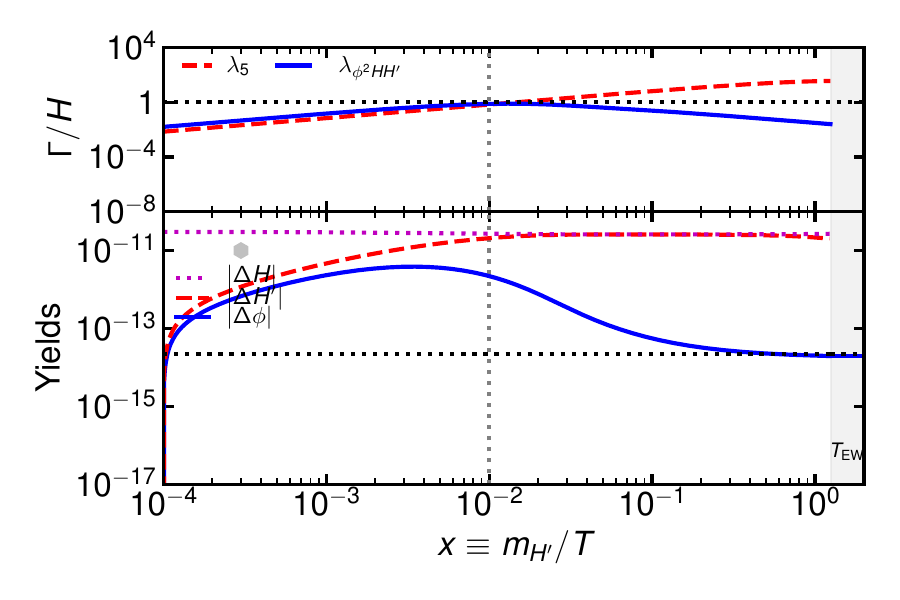}  \\
\end{tabular}
\vspace{-0.3cm}
\caption{Evolution of the asymmetries in $H$, $H'$ and $\phi$ as a function of temperature for the case with $m_{H'} = 200\,{\rm GeV}$ and $\delta m_{H^0} = 450\,{\rm keV} $, i.e. $\lambda_5 =3\times 10^{-6}$, see Eq.~\eqref{eq:lambda5DDcondition}. The top panel shows the region of parameter space where the observed dark matter abundance is obtained in blue. The other panels show the temperature evolution corresponding to each benchmark. The upper sub-panel shows the interaction rates for the various processes, in particular those equilibrating the Higgs doublets ($\lambda_5$) and the scalar $(\lambda_{\phi^2 HH'})$. The vertical dotted lines indicate $T = m_\phi$. } \label{fig:Yield}
\end{figure*}

\subsection*{Parameter Space Exploration}

In Fig.~\ref{fig:Yield} we show some examples of the evolution of the various asymmetries within our scenario. As highlighted in the text, we are working in a regime where the $\lambda_5$ interactions always thermalize and this is clearly seen by $\Delta H'$ equilibrating with $\Delta H$. What changes significantly in these plots is the timing and maximal strength of the rates induced by $\lambda_{\phi^2 H H'}$. There we clearly see that the right dark matter abundance is obtained either in the almost freeze-in regime (see e.g. the light green star or purple dot cases and discussion in the main text) or from larger rate situations, leading to a larger freeze-in production of the $\phi$ asymmetry, that is subsequently depleted (see e.g. the dark green cross and silver hexagon cases). A smaller value (red triangle) or larger value (cyan triangle) of $m_\phi$  leads to less and more asymmetry respectively. See the main text for more explanations.

In more detail,  when the temperature goes down and approaches the value of $m_\phi$, $\phi$ becomes non-relativistic. If by this time the $\lambda_5$ rate was still far from thermal equilibrium, the dynamical way to account for the observed relic density would be an ``ordinary" freeze-in one: the asymmetry would slowly increase out-of-equilibrium until it freezes at $T\sim m_\phi$ when the production becomes Boltzmann suppressed. In this case there would be only one value of $\lambda_{\phi^2HH'}$ that would lead to the observed relic density. However, the direct detection lower limit on the  $\delta m_{H_0}$ splitting implies that the $\lambda_5$ rate enters thermal equilibrium at a temperature above (or at least of order of) $m_\phi$. This results in a suppression of the asymmetry production, or even a depletion of the asymmetry, so that the dynamical way to obtain in our scenario the observed relic density is in fact not a pure freeze-in one.

The key feature  is that once the $\lambda_5$ rate is about to be in thermal equilibrium, 
the chemical potentials of $H$ and $H'$ become approximately equal, $\mu_{H'}\simeq \mu_H$  (i.e. $\Delta H'\simeq \Delta H$).  This means that from this time and until $\phi$ gets non-relativistic, the $\lambda_{\phi^2HH'}$ interaction, instead of creating an asymmetry, tends to erase it, see for instance Eq.~(\ref{eq:simple_phi_equation}). This is due to the fact that this interaction tends towards a situation where $2 \mu_\phi=\mu_{H}-\mu_{H'}$. Thus the sooner the  $\lambda_5$  enters into thermal equilibrium (i.e. the larger the $\lambda_5$ coupling and corresponding mass splitting), and/or the larger $\lambda_{\phi^2HH'}$ is, the larger is the depletion.

In practice the above implies that for fixed values of the other parameters, the observed relic density can be obtained for two different values of $\lambda_{\phi^2HH'}$, see the two branches in 
Fig.~\ref{fig:Mechanism_results}. The lower branch corresponds to a situation where this interaction is far from reaching thermal equilibrium, so that the asymmetry slowly grows until $\lambda_5$ is about to be in thermal equilibrium, at which point the asymmetry gets slowly suppressed, or grows less. This is the situation of the benchmark point in the right panel of Fig.~\ref{fig:Mechanism_results}. Values of $\lambda_{\phi^2HH'}$ below (above) this branch give too little (too much) freeze-in production. However, if one increases this interaction further, at some point it is large enough to be not far from thermal equilibrium at $T\sim m_\phi$, and the  $2 \mu_\phi\simeq \mu_{H}-\mu_{H'}\simeq 0$ relation holds, which means that the final asymmetry produced quickly drops. This leads to the upper branch in 
Fig.~\ref{fig:Mechanism_results}.

\begin{figure*}[!t]
\centering
 \hspace{-0.3cm} \includegraphics[width=0.45\textwidth]{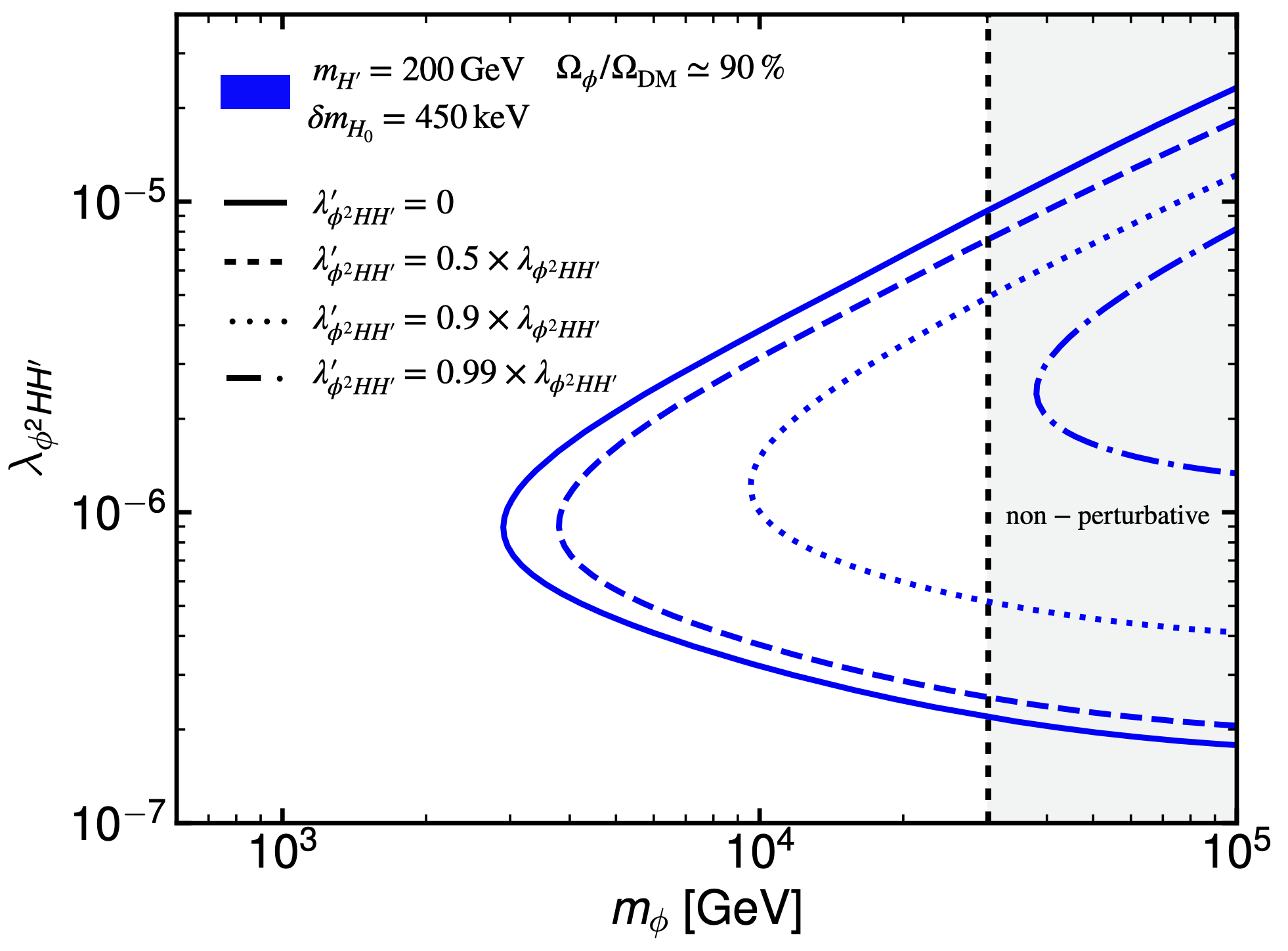} \hspace{-0.1cm} \includegraphics[width=0.45\textwidth]{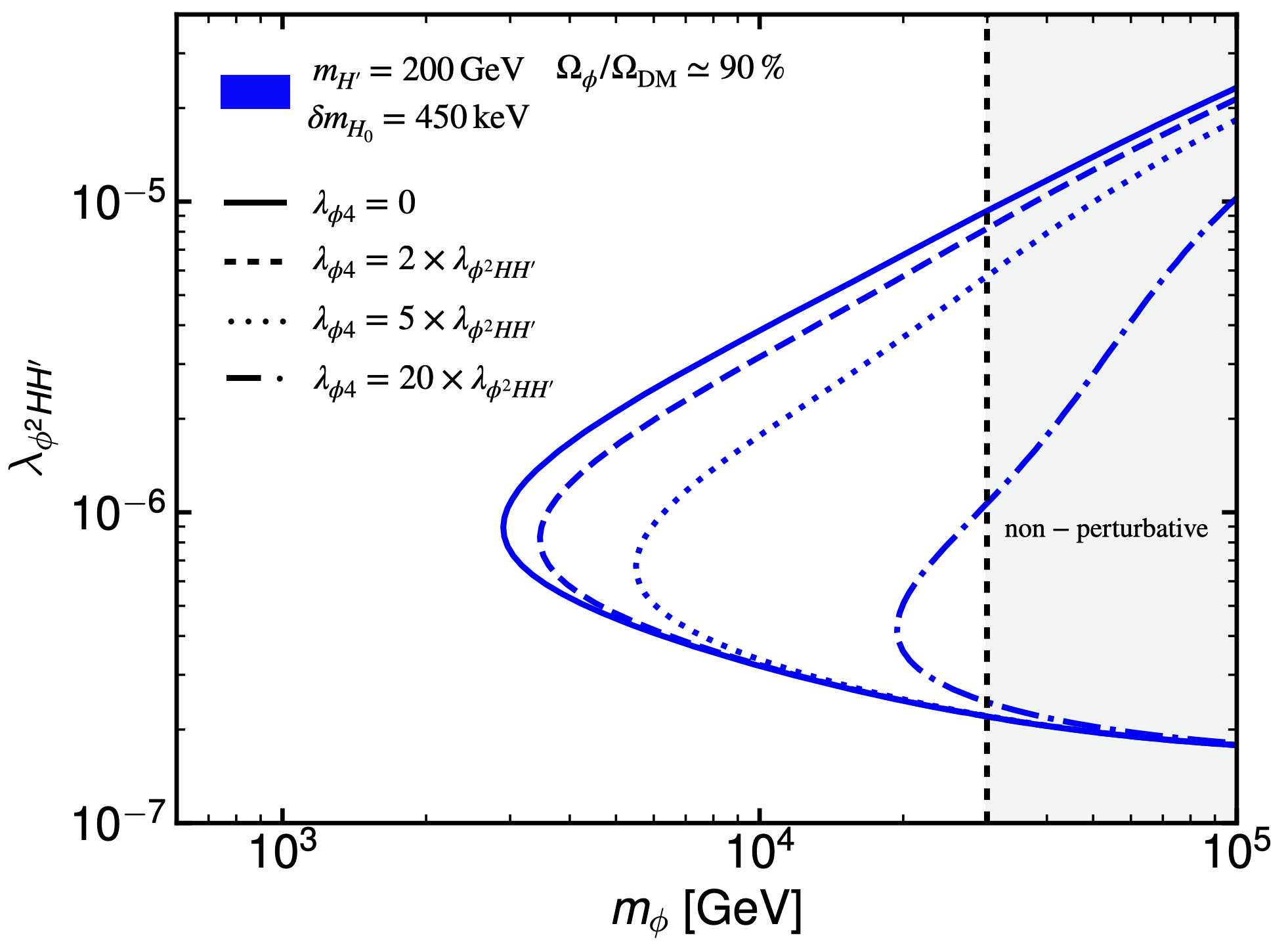} \hspace{-0.1cm}  
\vspace{-0.2cm}
\caption{\modific{Parameter space in the $\lambda_{\phi^2HH'}$-$m_\phi$ plane for a fixed value of $m_{H'} = 200\,{\rm GeV}$ and $\delta m_{H^0} = 450\,{\rm keV} $, i.e. $\lambda_5 =3\times 10^{-6}$, for various values of the $\lambda_{\phi^2HH'}'$ coupling (left) and for $\lambda_{\phi_4}$ (right). Unless $|\lambda_{\phi^2HH'}'|$ is finely tuned around $|\lambda_{\phi^2HH'}|$ or $|\lambda_{\phi 4}|\gg |\lambda_{\phi^2HH'}|$ the parameter space remains effectively unaltered with respect to the case $\lambda_{\phi^2 HH'}' = \lambda_{\phi 4} = 0$.} 
}
\label{fig:Impactothercouplings}
\end{figure*}

\modific{\textbf{Impact of additional asymmetric quartics:} In the main body we have ignored two couplings which are perfectly compatible with the $Z_4$ symmetry of our model: $\lambda_{\phi^2HH'}'\phi^2  H H'^\dagger/2+{\rm h.c.}$ and $\lambda_{\phi4}(\phi^4+{\rm h.c.})/4!$ in addition to $\lambda_{\phi^2HH'}/2\, \phi^2  H^\dagger H' +{\rm h.c.}$ and $\lambda_5/2\,(H^\dagger H')^2+{\rm h.c.}$. In Fig.~\ref{fig:Impactothercouplings} we showcase the impact that these couplings have for the parameter space of interest. From the left panel we can see that unless $\lambda_{\phi^2HH'}'$ is very similar to $\lambda_{\phi^2HH'}$ the parameter space remains effectively unaltered. In the limit in which $|\lambda_{\phi^2HH'}'| \to |\lambda_{\phi^2HH'}|$ no asymmetry in $\phi$ can be developed and hence the efficiency of the mechanism diminishes. The right panel shows the impact of turning on $\lambda_{\phi 4}$. One can observe that this impact is actually rather minor, unless it is significantly larger than $\lambda_{\phi^2 HH'}$. The fact that both of these couplings have a small impact allows us, for simplicity, to neglect them in the main body of the paper.} 

\end{document}